\documentclass[twocolumn]{aastex6}
\usepackage{hyperref}
\usepackage{amsmath}
\usepackage{graphicx, subfigure}

\shorttitle{Accretion, Feedback and Dynamics of BHs in CNDs}
\shortauthors{Souza Lima et al.}

\begin{document}

\title{The Pairing of Accreting Massive Black Holes in Multiphase Circumnuclear Disks: \\the interplay between radiative cooling, star formation, and feedback processes}

\author{Rafael Souza Lima\altaffilmark{1},
	    Lucio Mayer\altaffilmark{1},
	    Pedro R. Capelo\altaffilmark{1}, and
	    Jillian M. Bellovary\altaffilmark{2}
	    }
\email{rafael@physik.uzh.ch}
\altaffiltext{1}{Center for Theoretical Astrophysics and Cosmology, Institute for Computational Science, University of Zurich, Winterthurerstrasse 190, CH-8057 Z\"{u}rich, Switzerland}
\altaffiltext{2}{Department of Physics, Queensborough Community College, 222-05 56th Avenue, Bayside, NY 11364, USA}

\begin{abstract}
{We study the orbital decay of a pair of massive black holes (BHs) with masses $5 \times 10^5$ and $10^7$~M$_{\odot}$, using hydrodynamical simulations of circumnuclear disks (CNDs) with the alternating presence of sub-grid physics such as radiative cooling, star formation, supernova feedback, BH accretion and feedback. In the absence of such processes, the orbit of the secondary BH decays over timescales of ${\sim}10$~Myr to the center of the CND, where the primary BH resides. When strong dissipation operates in CNDs, fragmentation into massive objects the size of giant molecular clouds and with densities in the range $10^{4}$--$10^{7}$~amu~cm$^{-3}$ occurs, causing stochastic torques and hits that can eject the secondary BH from the midplane. Outside the plane, the low-density medium provides only weak drag, and the BH return is governed by inefficient dynamical friction. In rare cases, clump-BH interactions can lead to a faster decay. Feedback processes lead to outflows, but do not change significantly the overall density of the CND midplane. However, with a spherically distributed BH feedback a hot bubble is generated behind the secondary, which almost shuts off dynamical friction, a phenomenon we dub ``wake evacuation'', leading to delays in the decay of possibly ${\sim}0.3$~Gyr. We discuss the non-trivial implications on the discovery space of the eLISA telescope. Our results suggest the largest uncertainty in predicting BH merger rates lies in the potentially wide variety of galaxy host systems, with different degrees of gas dissipation and heating, yielding decay timescales from ${\sim}10$ to ${\sim}300$~Myr.}
\end{abstract}



\section{Introduction}

In a wide range of galaxy merger remnants, two or more massive black holes (BHs) are expected. This is mainly due to two reasons. First, there is compelling evidence for the existence of massive BHs in the center of massive galaxies \citep{richstone98, magorrian98}. Second, in the $\Lambda$CDM model, the assembly of larger galaxies happens through hierarchical merging \citep{white78}. Therefore, if more than one galaxy undergoing a given merging process hosts a BH, then two or more BHs are expected in the merger remnant. Previous studies of BH pair evolution in galaxy mergers suggest that BHs are driven to the center of the new host galaxy \citep{begelman80}, by losing orbital energy and angular momentum via processes such as dynamical friction against the gaseous and stellar background \citep{ostriker99,chandrasekhar43,binney08}. After the galaxies merge, the BHs reach a separation of less than $\sim$10~pc in relatively short timescales, compared to those in which the original galaxies merged \citep{callegari09,wassenhove14,capelo15}.

At a given stage of a gas-rich galaxy merger, the BH pair is expected to reside in a gaseous circumnuclear disk (CND). The formation of CNDs is predicted by simulations of mergers \citep{mayer07, mayer10}, furthermore it is corroborated by observations of nearby merging galaxies and galaxies that have undergone recent mergers \citep{medling14}. The mechanisms that compress and drive gas inflows to the nuclear regions of a merger \citep{mayer07, mayer10, capelo16} likely also induce star formation in some CNDs \citep{downes98}. This idea is supported by several reports of massive star formation activity in the nuclei of galaxy mergers. CNDs with strong star formation activity are expected to be clumpy \citep{klessen10, agertz09a, tasker09}, since sites of star formation are gravitationally unstable regions of gas, which can be associated to relatively cold and overdense gas clouds.

Recent studies on BH pair dynamics in CNDs attempted to track the influence of different interstellar medium properties, such as density contrasts (smooth vs. inhomogeneous), star formation, and gas accretion onto the BHs \citep{escala05, dotti07, mayer13, fiacconi13, roskar15, delvalle15, lupi15}. The effects of a clumpy environment on the dynamics of a BH pair in a range of BH mass ratios (1:1 to 1:5) in a gaseous CND are studied in \citet{fiacconi13}. The ``lumpiness'' of the disk is triggered and tuned by considering an artificial cooling term in the energy equation satisfied by the gas particles. It was found that the orbit of the light secondary BH can be affected in a stochastic manner, often extending the decay time in comparison to the smooth disk scenario. 

A similar result was obtained in \citet{roskar15} while considering an extended set of physical processes, however in galactic-scale simulations. According to this study, these processes affect the nature of the ambient in which the orbital decay process takes place. An implementation of a cooling term that accounts for energy-dissipation rates derived in atomic physics was considered, in contrast to the work of \citet{fiacconi13}. 

In the present work, the evolution of BH pairs in CNDs is studied. The effects of gas cooling, star formation, and SN feedback on the interstellar medium are included, using sub-grid prescriptions drawn from realistic simulations of galaxy formation that were not employed yet in simulations of CNDs (the gravitational force being resolved down to 0.5~pc), except in the work of \citet{roskar15}. For the first time, the effects of BH accretion and feedback from active galactic nuclei (AGN) in simulations of CNDs are also considered. This study represents a complementary step in the exploration initiated by \citet{roskar15}, by considering a large suite of CND runs rather than only a couple of galaxy merger runs. In this way we aim to understand in detail, and systematically, the effects of the various physical processes on the orbital decay. The BH masses are chosen to be in the range of those relevant for the future  evolved Laser Interferometer Space Antenna (eLISA). Concerning other previous similar works, \citet{fiacconi13} carried out a large set of CND simulations with massive BH pairs, but did not consider the effects of star formation and SNe, nor BH accretion or AGN feedback. The work of \citet{delvalle15}, which investigated the dynamics of BH pairs in star-forming CNDs, supports the picture that the orbits can be stochastically perturbed by high-density gas clumps, although they also find that the strength of such interactions is sensitive to the BHs softening length, decaying when the softening decreases. This was an important feature for the closest BH-clump encounters that were marginally resolved, at separations smaller than the gravitational softening assigned to the gas particles. In our case deflections of the BH orbit by gravitational scattering are seen to occur at clump-BH separations larger than the softening length, and also larger than the BH sphere of influence, hence we believe clump-BH interactions are well resolved. Besides, in their implementation of SN feedback,  they let the gas cool after deposition of the energy of SN explosions into the surrounding gas, which is known to lead to inefficient feedback, likely missing the effect of blastwaves. They also did not include BH accretion nor its feedback.

We thus ran a comprehensive set of simulations to disentangle the effects of different processes on the dynamics of a BH pair in a CND. The set up of the initial conditions of these simulations is described in \S\ref{sec:methods}, in addition to the implementations of the physics considered. \S\ref{sec:results} highlights the results of the individual influence of gas cooling, star formation rates, SN feedback and AGN feedback on the orbital decay of the BH pair. Finally, in \S\ref{sec:conclusions} we discuss the implications of the results on the current understanding of the process of orbital decay and on detection rates of future gravitational-wave experiments, the conclusions that can be drawn from them and limitations of the present work.


\section{Methods}\label{sec:methods}

Numerical simulations are performed using the Tree Smoothed Particle Hydrodynamics (SPH) $N$-body code \textsc{Gasoline} \citep{wadsley04}. The system under study comprises a CND, initially in hydrostatic equilibrium, with two embedded BHs initially separated by a distance of the order of the disk's scale radius. We start by describing the initial conditions setup and then continue with a description of the different physics implemented in the simulations.


\subsection{Initial Conditions}

We used an initial disk model similar to the ``smooth set'' from \citet{fiacconi13}, therefore comparable to \citet{escala05} and \citet{dotti07}. The CND structural parameters are set to be commensurate to those of CNDs resulting from galaxy merger simulations \citep{mayer07, roskar15} and observed in gas-rich interacting galaxies \citep{medling14}. As a reference, the typical range of molecular gas masses, which is believed to be the dominant gas phase by mass in CNDs, is 10$^{8-9}$~M$_{\odot}$ for $\sim$100~pc scale CNDs  \citep{izumi+16}. The gaseous disk is represented by $2 \times 10^5$ SPH (gas) particles, with total mass $M_{\textrm{d}} = 10^8$~M$_{\odot}$, and follows a \citet{mestel1963} surface density profile with scale radius $R_{\textrm{d}} = 100$~pc and maximum radius $\sim$150~pc, with a Gaussian vertical structure of scale height $z_{\textrm{d}} = h R_{\textrm{d}}$, where $h=0.05$ is the aspect ratio. The initial temperature of all the gas particles is set to $T_0 = 2 \times 10^4$ K and the initial equilibrium assumes a polytropic equation of state with index $\gamma = 1.4$. The disk is embedded in a concentric \citet{plummer1911} spheroid that models the inner part of a galactic bulge, represented by $10^6$ dark particles, with total mass $M_{\rm P} = 5 \times 10^8$~M$_{\odot}$, scale radius $r_{\rm P} = 50$~pc, and maximum radius $\sim$500~pc.
The primary BH, of initial mass $M_{01} = 10^7$~M$_{\odot}$, is placed at the center of the CND, then the system is left to relax towards equilibrium for 10~Myr. The secondary BH, of initial mass $M_{02} = M_{01} / 20$, is positioned at 80~pc separation from the disk's longitudinal axis and slightly displaced from the midplane at a height of about 3~pc. Its initial radial and tangential velocity components are both equal to approximately 110~km~s$^{-1}$, corresponding to a total initial speed about the same as the local circular velocity and enforcing an eccentric orbit (eccentricity greater than 0.5). The choice of such an eccentric orbit is motivated by the results of pc-resolution galaxy merger simulations modeling the formation of the CND \citep[][]{mayer07, roskar15}. The force resolution is determined by a gravitational softening $\epsilon_{\rm F} = 0.5$~pc common to all particles.


\subsection{Radiative cooling, star formation, and supernova feedback}

All runs use the same initial configuration of a CND with two embedded BHs, but we greatly vary the sub-grid physics implemented in each simulation (see Table~\ref{tab:runs}). In all simulations, we solve the internal energy equation using the standard density-energy formulation of SPH with a standard Monaghan artificial viscosity \citep{wadsley04} and set the adiabatic index to $\gamma=1.4$, unless otherwise specified (a lower $\gamma$ leads to a softer equation of state and hence to less compressional/shock heating, which can be an important effect in self-gravitating disks; see e.g. \citealt{mayer2004}). We include Compton and non-equilibrium atomic cooling rates for a gas composed of only hydrogen and helium \citep{wadsley04}. While metal-line cooling may play a role in the metal-enriched regions at the center of galaxy merger remnants, where CNDs arise, various heating sources would matter as well in the high-star-formation environment expected in such CNDs. These include radiative and photoionization feedback from stellar radiation of OB stars, radiation pressure, photoelectric heating of dust grains and so on, which, in combination with high absorption and scattering opacities in the dense gas, would lead to mean gas temperatures in excess of 1000~K outside cold star-forming clumps, as shown in \citet{roskar15}, where many of these effects are included. Since our focus here is to explore a relatively large set of simulations, we prefer to limit the numerical complexity of individual simulations and rely on the fact that, with the high resolution of our simulations (below pc scales) and high mean densities of CNDs, the formation of a multiphase medium with dense clumps will occur even with just cooling via recombination and radiative transitions of hydrogen and helium.

Star formation is implemented according to \citet{stinson06}, therefore using a Schmidt law \citep{schmidt59,schmidt63} and stochastic conversion of gas particles into star particles for the coldest and densest regions of gas. The star formation efficiency parameter is set to 0.05 in the Schmidt law. The gas minimum-density threshold for star formation is $\rho_{\rm thresh} = 10^5$ amu cm$^{-3}$ and the maximum-temperature threshold, $T_{\rm SF}$, is determined empirically by constraining the global star formation efficiency, i.e. by requiring that the total mass of stars formed over the initial total mass of gas should not exceed 10\% over the typical duration of the simulations, consistent with the observations of starbursts in merging galaxies (\citealt{medling14}; note that starburst systems appear to have a slightly higher star formation efficiency relative to normal star-forming galaxies where this is closer to a few percent on a range of scales). The latter is the star formation efficiency resulting when the self-regulating effect of feedback is also included (hence in runs without feedback this can be exceeded significantly over long timescales - see Table~\ref{tab:mass_ratios}). This sets $T_{\rm SF} = 4500$~K. While this may seem a large temperature threshold, we recall that we do not use metal-line cooling, hence gas will not cool below a few thousand~K.

SN feedback also follows the blastwave recipe from \citet{stinson06}, based on the analytical  model described in \citet{mckee77}. Blastwave shocks are assumed to convert kinetic energy into thermal energy on scales smaller than the simulation resolution, therefore only their thermal effects on the surrounding gas are accounted for. Each SN deposits a net energy of $\epsilon_{\rm SN} \times 10^{51}$~erg in the form of heat into the neighboring gas particles (within the SPH kernel), and we adopt a feedback efficiency  $\epsilon_{\rm SN} = 0.8$ \citep{guedes11a}. Cooling is suppressed for particles within a radius computed via the blastwave solution of \citet{mckee77} and for a timescale of a few 10~Myr, which corresponds to the snowplow phase. This ``delayed cooling'' recipe, which has been shown to reproduce realistic galaxies across a wide range of masses \citep[e.g.][]{guedes11b,Shen_et_al_2014,Sokolowska_et_al_2016a,Sokolowska_et_al_2016b}, applies only to SNe Type II, while for SNe Type I only a simple thermal energy dump within the local  SPH kernel is applied. With the conservative parameters adopted in this paper, our sub-grid model can overestimate the star formation efficiency in the dense ISM of galaxies at high redshift \citep[][]{Sokolowska_et_al_2016a}, but this is not a concern here as our models are supposed to represent low-redshift CNDs.


\subsection{Accretion onto black holes}

An analytical expression that estimates the hydrodynamic spherical accretion rate onto a non-rotating BH for an infinite mass distribution of gas with no angular momentum is obtained in \citet{bondi52} \citep[see also][]{hoyle39,bondi44}. The accretion rate $\dot{M}_{\rm Bondi}$ onto a BH of mass $M_{\rm BH}$ is expressed in terms of the ambient density $\rho_{\rm gas}$, the sound speed $c_{\rm s}$ and the relative velocity $V$ of the gas with respect to the massive object via the Bondi--Hoyle--Lyttleton (hereafter Bondi) equation:

\begin{equation}
\label{eq:accretion}
\dot{M}_{\rm Bondi} = \lambda 4 \pi G^{2} M_{\rm BH}^{2}\frac{\rho_{\rm gas}}{\left( V^{2} + c_{\rm s}^{2}\right)^{3/2}},
\end{equation}

\noindent where $\lambda$ is a constant of order unity determined by the equation of state of the gas, and $G$ is the gravitational constant. In simulations, prescriptions based on Equation \ref{eq:accretion} can lead to overestimates or underestimates of the accretion, depending on how the variables above are calculated and on the resolution \citep[e.g.][]{negri16}. However, despite the adoption of a boost factor equal to unity, we suspect that, if anything, it is the uncertainty in AGN feedback that plays the most important role. On the other end, accretion recipes that account for the angular momentum of the gas, such as those based on gravitational torques \citep{hopkins12, alcazar17}, could lead to lower accretion rates. This must be explored in future papers. Yet we recall that the overall scenario presented here, in which perturbations onto the orbit of the BH, including ejections, are the main player, seems to hold irrespective of the inclusion, or not, of BH accretion and feedback. Our numerical implementation of accretion \citep{bellovary10} considers the parameters in Equation (\ref{eq:accretion}) computed within one smoothing length from the BH. In this case, $\lambda$ is simply a normalization factor, which is intended to reproduce phenomenologically the accretion rates estimated from observational evidence. While in the past large values of $\lambda$ have been used to compensate for the lack of dense gas due to the low resolution of simulations \citep[e.g.][]{dimatteo05}, the sub-pc resolution of our simulations allows us to  set  $\lambda = 1$ \citep[see also][]{callegari09,tamburello2016}. The accretion rate is capped at the Eddington limit. This critical rate, $\dot{M}_{\rm Edd}$, is proportional to the mass of the accreting object,

\begin{equation}
\dot{M}_{\rm Edd} = \frac{4 \pi G m_{\rm p}}{\epsilon_{\rm r} \sigma_{\rm T} c} M_{\rm BH} = \frac{1}{\tau_{\rm Salp}} M_{\rm BH},
\label{eq:salp acc}
\end{equation}

\noindent where $m_{\rm p}$ is the proton mass, $\epsilon_{\rm r}$ is an efficiency parameter between 0 and 1 representing the fraction of the mass-energy budget that is assumed to be radiated away (we use $\epsilon_{\rm r} = 0.1$), $\sigma_{\rm T}$ is the Thomson cross section, and $c$ is the speed of light in vacuum. The time $\tau_{\rm Salp} \approx 45$~Myr, known as the \citet{salpeter64} time, is a proportionality factor associated to a typical timescale during which the BH feeding at this rate grows by a factor of $e$. In the simulations, the gas particles that are eligible to be accreted by the BH will have their mass gradually transferred to the compact object according to Equation (\ref{eq:accretion}) whenever $\dot{M}_{\rm Bondi} < \dot{M}_{\rm Edd}$, or based on the rate computed with Equation (\ref{eq:salp acc}) otherwise.


\subsection{AGN feedback}

The emitted radiative power in the process of BH accretion (AGN feedback) is assumed to be proportional to the rate at which mass is accreted. A fraction $\epsilon_{\rm fb}$ of it heats up the gas around the BH \citep{bellovary10}. Therefore, the energy $E_{\rm fb}$ released via AGN feedback that thermally couples to the gas per time step $dt$ can be expressed as

\begin{equation}
\label{eq:feedback}
E_{\rm fb} = \epsilon_{\rm fb} \epsilon_{\rm r} \dot{M}_{\rm gas} c^2 dt.
\end{equation}

In the simulations including AGN feedback, part of the energy released by the accreting BH is thermally coupled to the neighboring gas particles according to Equation (\ref{eq:feedback}), with $\epsilon_{\rm fb} = 0.05$ \citep{dimatteo05, springel05b,johansson09,bonoli16}.


\section{Results}\label{sec:results}


\subsection{Orbital decay}\label{sec:orbital decay}

We have performed a small set of adiabatic\footnote{In this paper, by ``adiabatic'' we mean runs in which we solve the internal energy equation with cooling switched off.} runs followed by the main set of simulations with radiative cooling, star formation and varying feedback processes. Table~\ref{tab:runs} describes the simulations used for the analysis appearing in this paper and defines the labels used throughout the text. The BH separation over time is plotted for some representative cases in Figure~\ref{fig:separation}. In the adiabatic runs, the decay occurs on a shorter timescale than that of all the other simulations (except in one case, see below run CSFSN). These runs, whose general behaviour is already well documented in the literature, serve as a benchmark for the subsequent simulations. The main common feature of all the runs that include radiative cooling, irrespective of the presence of other processes, is disk fragmentation. The latter produces a highly inhomogeneous, multiphase background, affecting the orbital decay by making it stochastic, and often leading to a much longer decay timescale (Figure~\ref{fig:separation}). This general outcome confirms the findings of \citet{fiacconi13}, who used a simple constant-cooling model and had no star formation, BH accretion, nor feedback processes implemented. The degree of clumpiness, however, depends on the specific physics implemented in the runs. In addition, the different physical processes, in particular SN and AGN feedback, 
affect the CND structure in other ways that can have a subsequent impact on the orbital decay, both globally and locally.

In the remainder we will describe in detail the results of the various runs, implementing an increasingly higher number of physical processes via sub-grid models, starting with a summary of the outcome in the idealized, but highly instructive case of adiabatic runs.


\subsubsection{Adiabatic runs}\label{sec:adiabatic}

\begin{figure}
\includegraphics[width=0.47\textwidth]{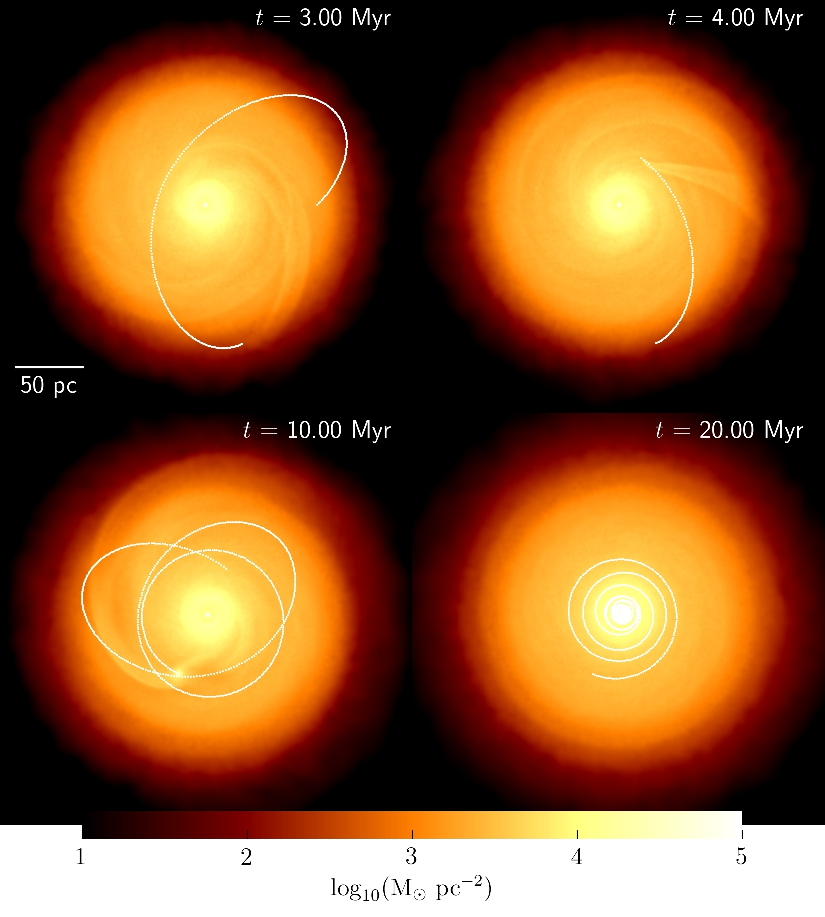}
\caption{Gas density face-on projection and secondary BH's trajectory (white lines) for the adiabatic run NAAD (see Table~\ref{tab:runs}) at different times. A scale bar on the top left panel shows the size in pc. The colour code represents the disk density integrated along the $z$-axis (perpendicular to the disk plane), which rotates counter-clockwise in this perspective. Trajectories are shown for time intervals between snapshots. On the top left panel, we see a leading spiral wake caused by the BH near an apocenter as a perturbation to the disk density. On the top right panel, a trailing wake is captured near a pericenter. The net effect of the wakes over a few orbits is to circularize the orbit. The lower left panel shows a resonant interaction between the BH and the gas as the orbit becomes circular. After circularizing, the gas extracts its angular momentum and the separation shrinks in a relatively short timescale, as clearly captured on the bottom right panel. The interplay between perturber and background is analogous to results found in the literature for planet migration.
\label{fig:naad}
}
\end{figure}

In the adiabatic runs, the disk remains smooth on all scales, and only the perturbation of the secondary BH as it travels through the disk is noticeable (Figure~\ref{fig:naad}).

\begin{figure}{
\includegraphics[width=0.45\textwidth]{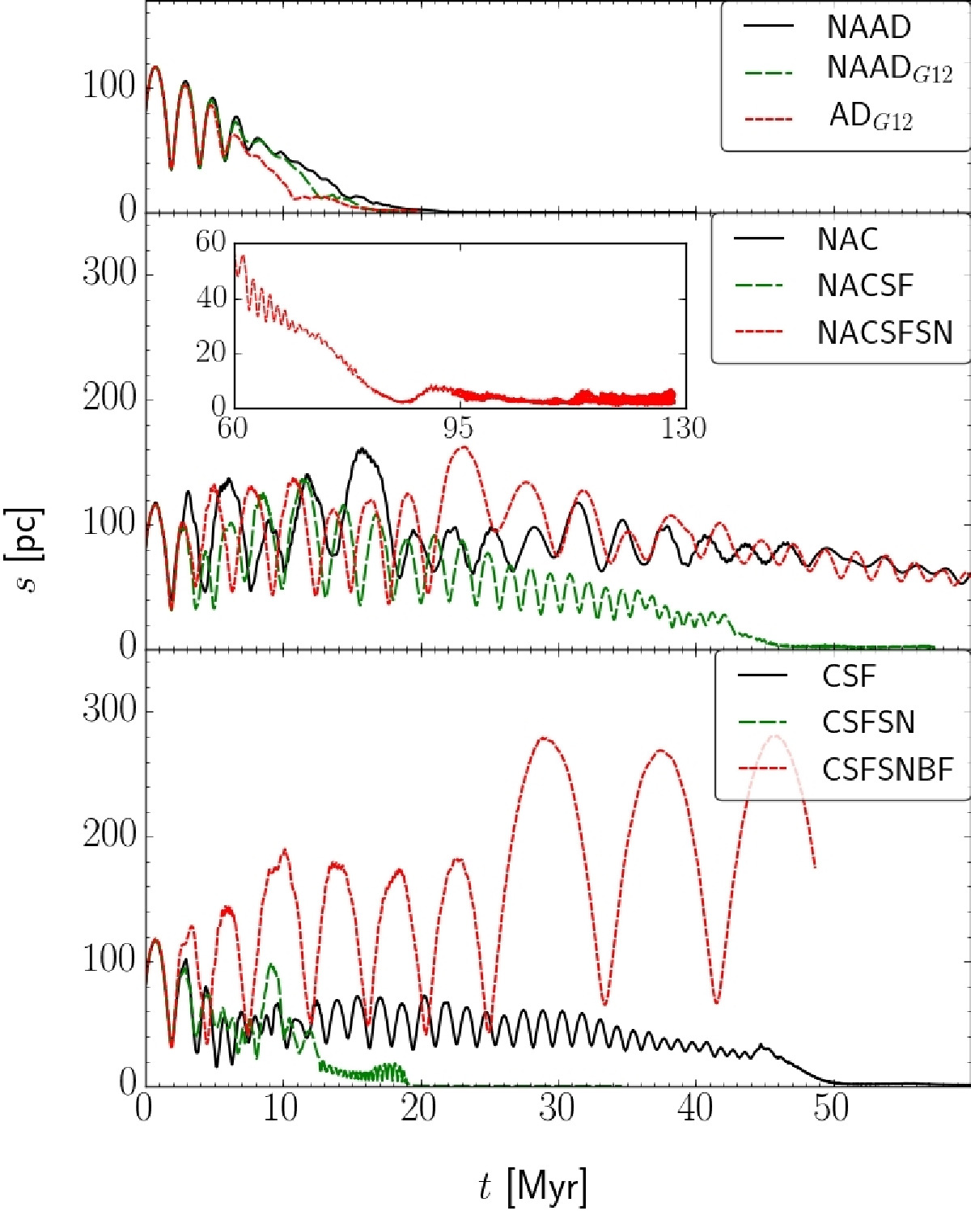}
\caption{Distance between the BHs for different runs (refer to Table~\ref{tab:runs}). The shortest timescales for the orbit of the secondary BH to fall under the one-pc scale separation from the central BH are observed in the adiabatic runs (top panel). Almost all the other runs that include gas cooling lead to prolonged, if present, orbital decay. The exception is run CSFSN (bottom panel), in which an encounter between the secondary BH and an overdense region of gas caused a relatively short decay time. Run CSFSNBF was stopped before 60~Myr due to computational limitations.}
\label{fig:separation}
}
\end{figure}

The BH separation falls to pc-scale in less than 20~Myr (see Figure~\ref{fig:separation}, top panel). During this time, the secondary BH acts as a perturber to the gas, inducing the formation of a wake,  and its orbit steadily circularizes (see Figure~\ref{fig:naad}), i.e. the apsides approach the same value, which is why the separation decreases at constant angular momentum (see Figure~\ref{fig:separation}, top panel; and Figure~\ref{fig:angular_momenta}). In this phase the orbital decay is well described by dynamical friction \citep{chandrasekhar43, ostriker99}.

\tabletypesize{\footnotesize}
\begin{deluxetable}{llllll}
\tablecolumns{5}
\tablecaption{Simulations overview.\label{tab:runs}}
\tablehead{
\colhead{Run label\tablenotemark{a}} & \colhead{Accretion} & \colhead{Cooling} & \colhead{Stars} & \colhead{SN} & \colhead{$\epsilon_{\rm fb}$}
}
\startdata
NAAD								& No		& No		& No		& No		& N.a.\\
NAAD$_{\rm G12}$\tablenotemark{b}		& No		& No		& No		& No		& N.a.\\
AD$_{\rm G12}$\tablenotemark{b}			& Yes	& No		& No		& No		& 0\\
NAC									& No		& Yes	& No		& No		& N.a.\\
NACSF								& No		& Yes	& Yes	& No		& N.a.\\
NACSFSN							& No		& Yes	& Yes	& Yes	& N.a.\\
C									& Yes	& Yes	& No		& No		& 0\\
CSF									& Yes	& Yes	& Yes	& No		& 0\\
CSFSN								& Yes	& Yes	& Yes	& Yes	& 0\\
CBF									& Yes	& Yes	& No		& No		& 0.05\\
CSFSNBF								& Yes	& Yes	& Yes	& Yes	& 0.05
\enddata
\tablenotetext{a}{ NA stands for \emph{no BH accretion}, AD for the \emph{adiabatic} runs, C for \emph{gas cooling}, SF for \emph{star formation}, SN for \emph{SN} feedback, and BF for \emph{feedback} due to accretion onto \emph{BHs}.}
\tablenotetext{b}{G12 stands for adiabatic/polytropic index 1.2 instead of 1.4.}
\end{deluxetable}

When gas accretion onto the BHs is considered, circularization happens in a slightly shorter time interval, driving the separation to less than 20~pc faster, although it does not change appreciably the time required to reach (sub-)pc scales, since this is dominated by a second decay phase.

After circularization occurs, indeed, dynamical friction by the background CND, as described in the conventional local-``wake'' approach of \citet{chandrasekhar43} and \citet{ostriker99}, becomes weak as the secondary BH has negligible velocity relative to the gaseous background (but not relative to the stellar bulge, which would still exert a drag). Instead, another regime ensues in which spiral density waves are triggered by the secondary BH and back-react by draining its orbital energy and angular momentum. This migration regime is analogous to Type I and Type III migration in planet formation and leads to a faster orbital decay compared to the dynamical friction phase \citep{mayer13}. This is characterized by fast extraction of angular momentum from the BH. The negative torques experienced by it have been shown to be associated with two mechanisms, namely the excitation of density waves in the disk near the Lindblad resonances, and the torques from the gas co-orbiting with the  BH. The latter co-orbital torques are stronger and cause fast inward migration, as highlighted by the fast steepening of the angular momentum evolution curve in Figure~\ref{fig:angular_momenta} (see \citealt{mayer13} for torque analysis in a similar case). The short timescale associated with this second stage of the migration process likely explains the absence of an evident gap-opening: the secondary BH simply does not have time to develop a gap in regions where the conditions would allow it, as it traverses those too quickly, as thoroughly studied and discussed in \citet{malik2015} for massive planets and brown dwarfs in protostellar disks.

\subsubsection{Effects of cooling}\label{sec:cooling}

\begin{figure}
\includegraphics[width=0.47\textwidth]{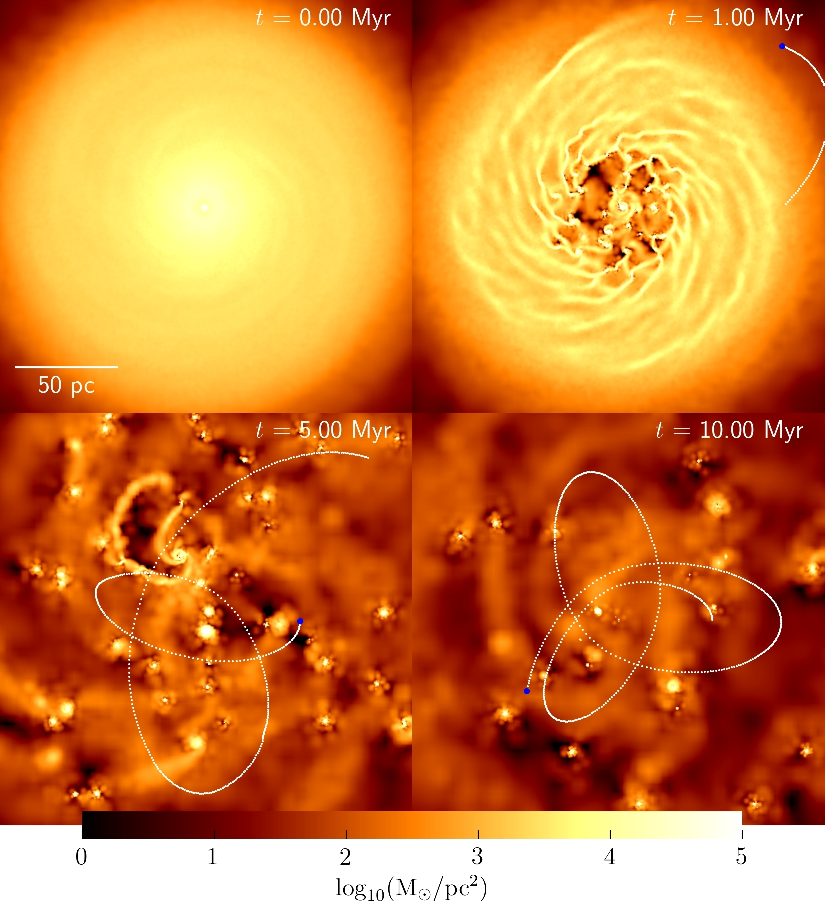}
\caption{Gas density face-on projection and secondary BH's trajectory for run C (see Table~\ref{tab:runs}), similar to Figure~\ref{fig:naad}. Notice the change of the scale and the surface density limits in contrast with that figure. The left and right upper panels show respectively the initial state of the gas and the fragmentation happening in the central regions of the disk. From the bottom left panel we notice that the fragmentation timescale is fairly smaller than the timescale for the secondary BH to complete one orbit. As time elapses, many gaseous clumps are formed and interact with each other, scattering and combining to form fewer larger clumps. The environmental differences along the BH path in comparison to the adiabatic runs are enormous. This leads to a completely different interplay between the gas disk and the BH, as the latter moves through a more contrasted density background.
\label{fig:c_face}
}
\end{figure}

\begin{figure}
\includegraphics[width=0.47\textwidth]{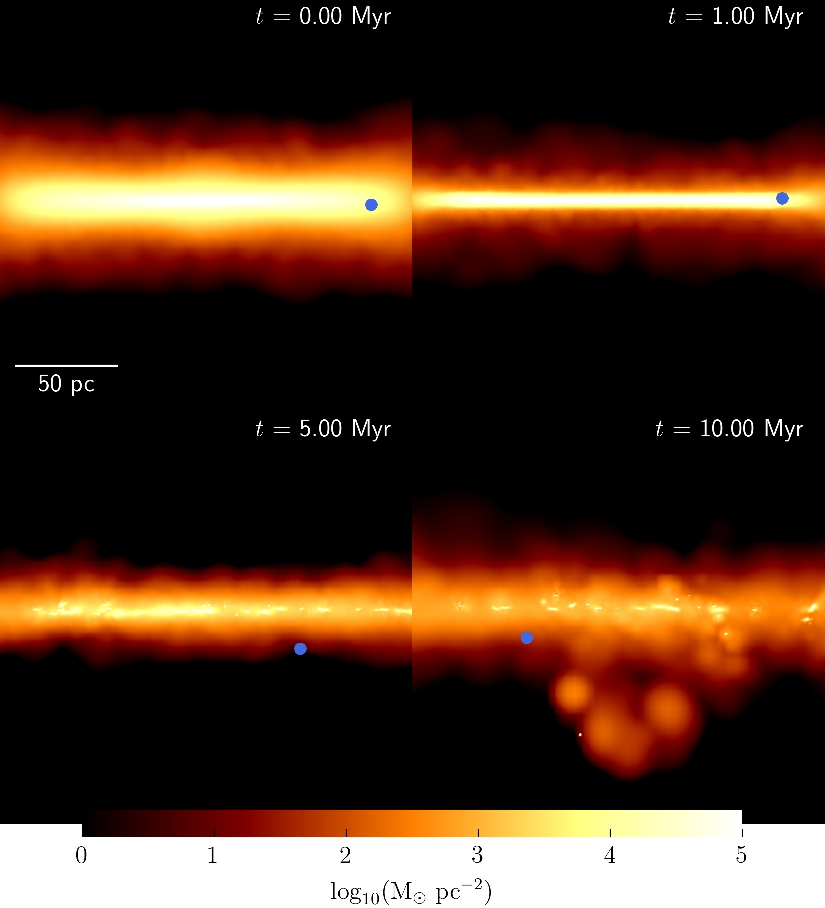}
\caption{Gas density edge-on projection and secondary BH's position for run C (see Table~\ref{tab:runs}). The first panel is the initial state of the disk and is also representative of the vertical structure of the disk in the adiabatic runs for all times. The blue dot indicates the secondary BH position. As the gas cools down, it becomes denser as it offers less pressure support against gravity, finally leading to formation of clumps that are scattered around. On the bottom right panel we can see a returning massive clump that was scattered out far from the disk plane.
\label{fig:c_side}
}
\end{figure}

\begin{figure}
\includegraphics[width=0.45\textwidth]{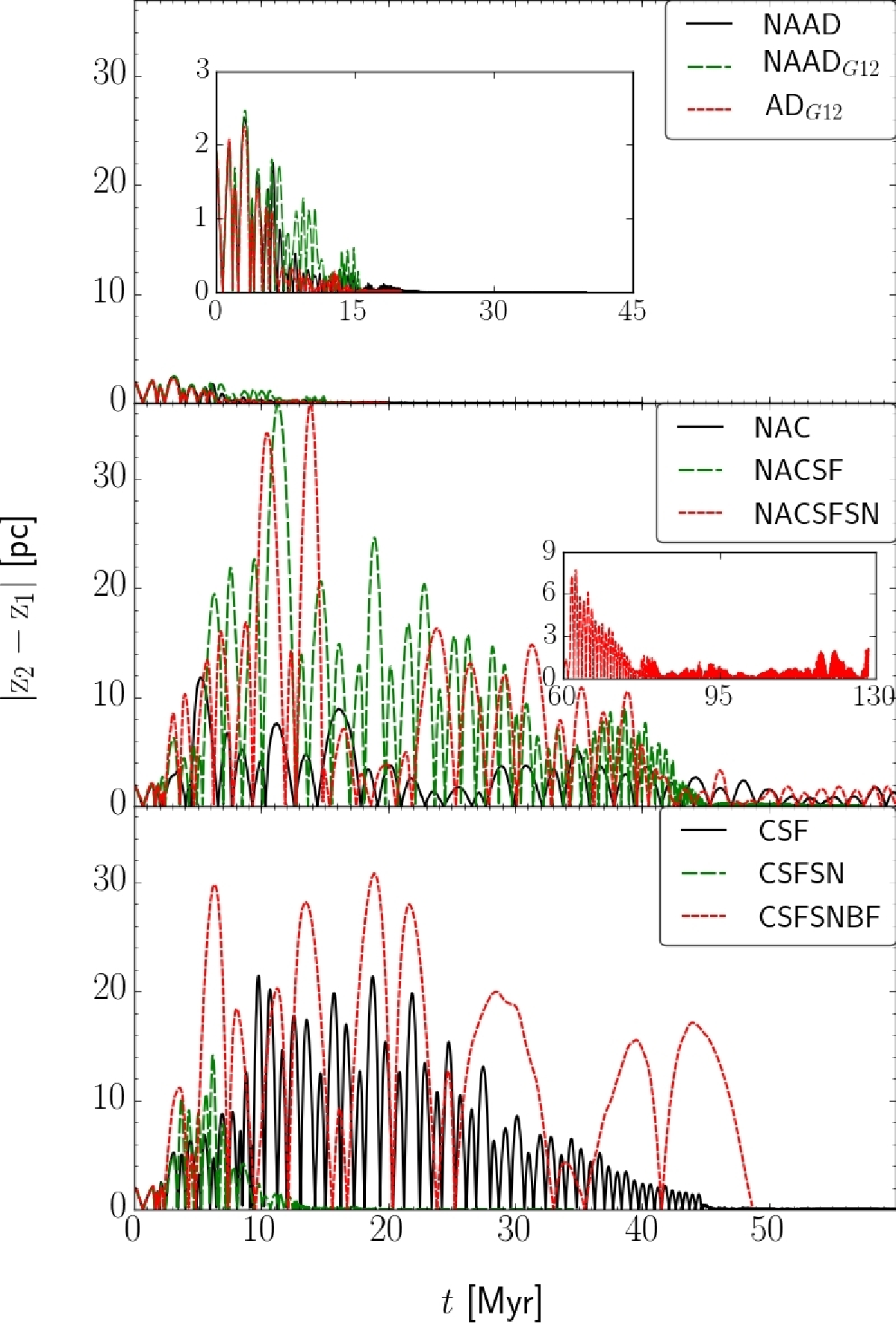}
\caption{Vertical separation between the BHs for different runs (refer to Table~\ref{tab:runs}). The plots indicate when ejections out of the disk plane occur. In the top panel (adiabatic runs), the secondary BH never exceeds the 5-pc (initial $z_{\rm d}$) range from the mid plane.
\label{fig:z_separation}
}
\end{figure}

We now proceed to study the outcome of the radiative-cooling runs. Cooling, even in absence of metals, is fast at the high densities of the CND, in particular it occurs on a timescale faster than the local orbital time (which is of order of 1~Myr at the half-mass radius of the disk). This leads to rapid fragmentation into clumps, as it is well studied and established in the literature on self-gravitating disks \citep[see e.g.][]{gammie2001,durisen2007}. The transition to a clumpy disk regime is abrupt in absence of any heating and/or star formation, taking place in only 0.3~Myr, namely in a fraction of the time it takes the secondary BH to complete the first orbit (see Figures~\ref{fig:c_face} and \ref{fig:c_side}). At this stage the disk enters a gravo-turbulent regime, in which collisions and tidal interactions between cold clumps occur as well as torquing by strong spiral density waves (Figure~\ref{fig:c_face}). Clumps typically gravitationally scatter each other but can also merge into heavier clumps, which can cause even stronger perturbations on the secondary BH. Ejection of the secondary BH from the disk plane can result, and indeed almost always occurs in our simulations (Figures~\ref{fig:c_side} and \ref{fig:z_separation}). The overall effect is to induce a stochastic behaviour of the orbital separation as opposed to the smooth behaviour in the non-fragmenting, adiabatic case (Figure~\ref{fig:separation}).

The diffuse background gas is also affected by torques, which produce a global  inward flux of mass and make the CND more centrally concentrated. This reflects the nature of self-gravity, which can be described by an effective viscosity acting throughout the disk even in absence of fragmentation \citep[][]{lin_pringle1987}.

In run NAC (see Table~\ref{tab:runs}), within 10~Myr, the inner 25-pc region goes from enclosing less than 20\% of the total mass of the disk and being smooth to having more than 50\% of the total disk mass in 3 clumps, one of about $3.19 \times 10^7$~M$_{\odot}$ enclosing the primary BH, another
 one of about $1.88 \times 10^7$~M$_{\odot}$ (heavier than the primary BH itself) at a distance of about 8~pc center to center 
(each have a radius larger than 1.5~pc), and the third one of about $6.7 \times 10^6$~M$_{\odot}$, still much heavier than the secondary 
BH, with respective densities ${\sim}1.5 \times 10^8$, $5 \times 10^7$, and $10^7$~amu cm$^{-3}$ (during this time, the secondary BH is always outside this inner region; see Figure~\ref{fig:separation}).

\begin{figure*}
	\subfigure[Distribution of clump masses]{
		\label{fig:histogram-mass}
		\includegraphics[width=0.99\textwidth]{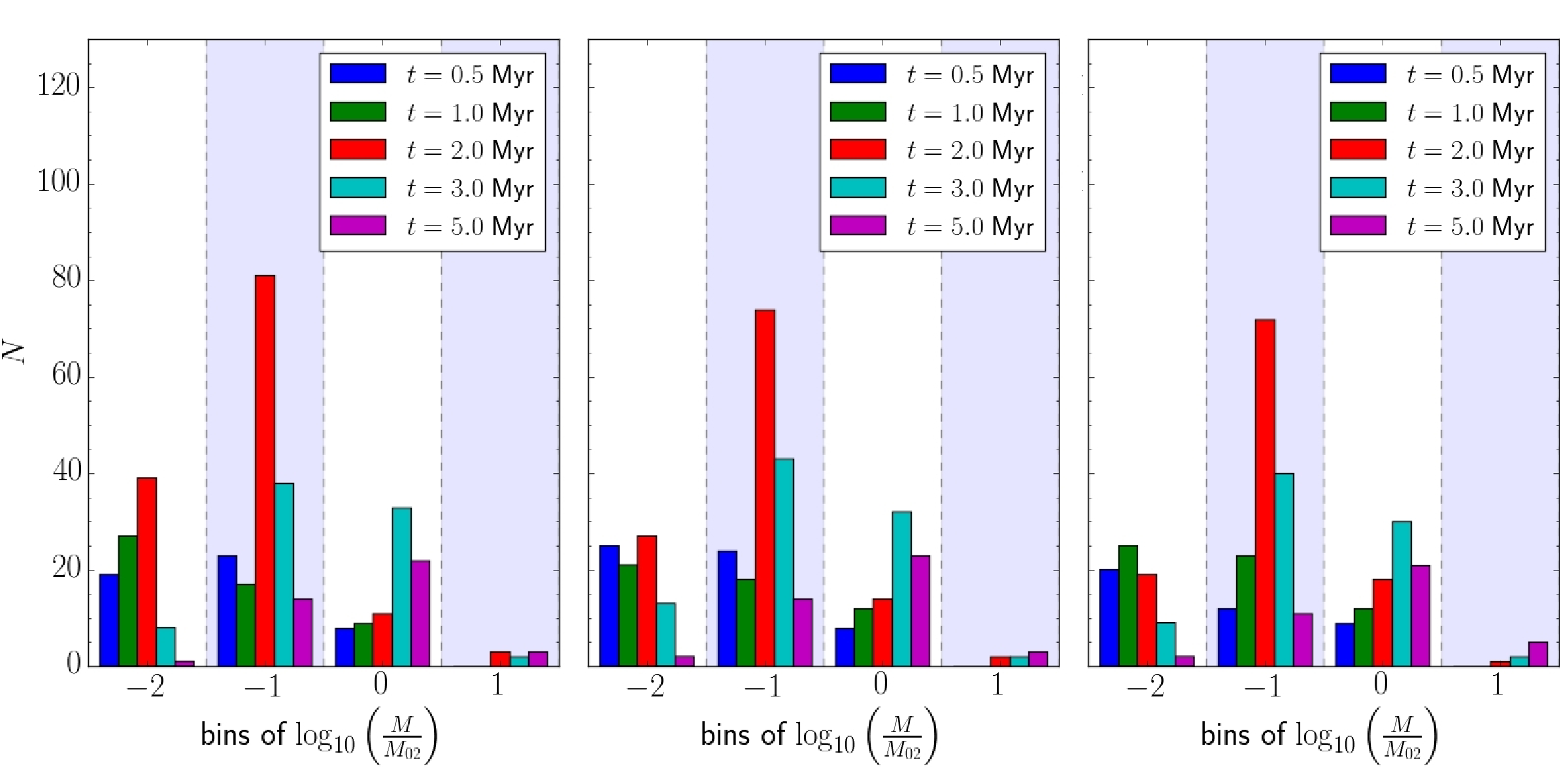}
	}\\
	\subfigure[Distribution of clump densities]{
		\label{fig:histogram-density}
		\includegraphics[width=0.99\textwidth]{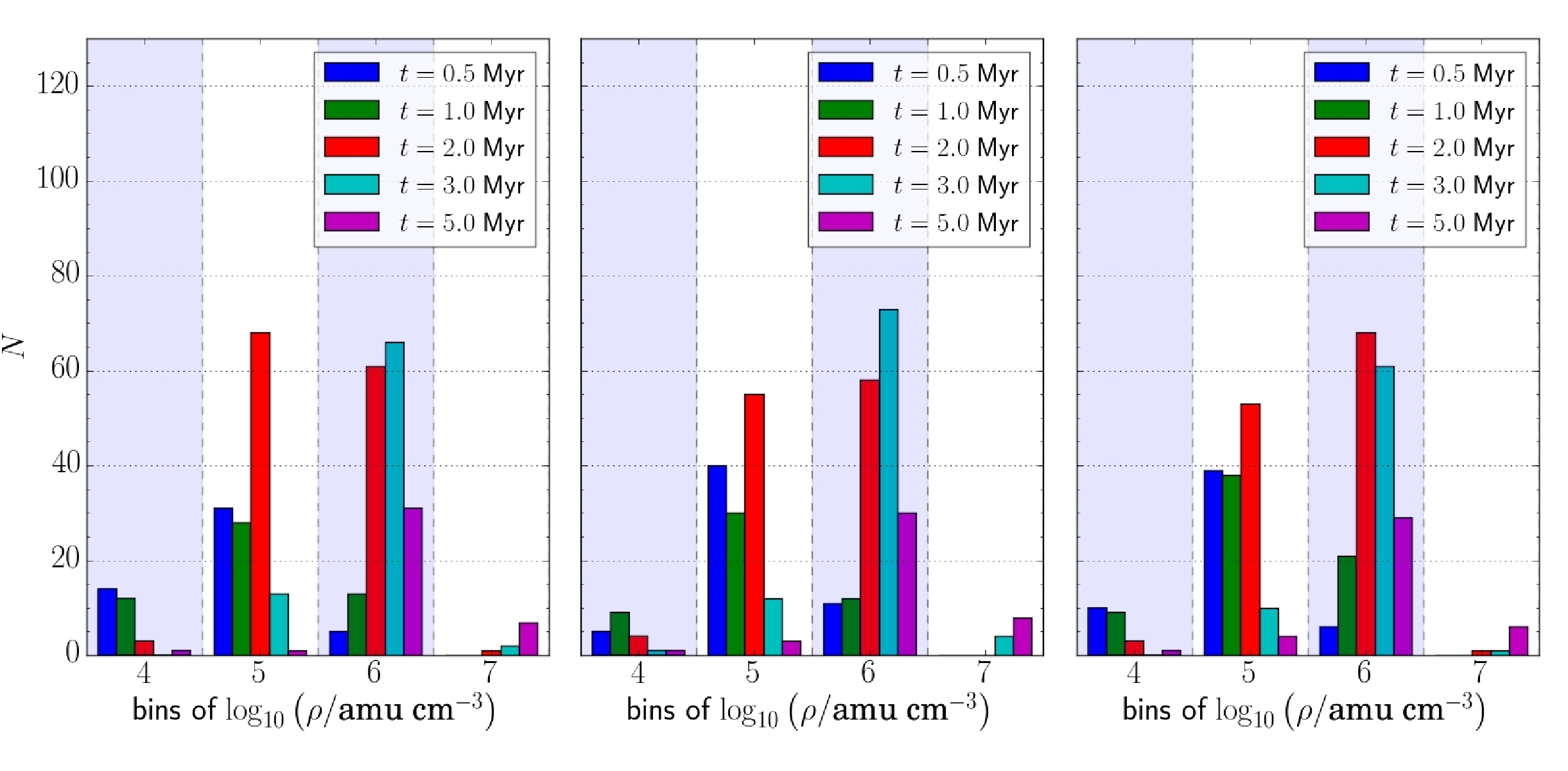}
	}
\caption{ \ref{fig:histogram-mass} Histogram of the clump mass distribution at different times for runs NAC (left), CSFSN (middle), and CSFSNBF (right). The horizontal axis is shown in terms of the initial mass of the secondary BH, to highlight the potential dynamical influence of a close encounter between the secondary BH and a clump in the disk. As time elapses, smaller clumps merge into larger ones, building up a sensible population of bound objects with mass comparable to or larger than the one of the secondary BH. For reference: $M_{02} = 5 \times 10^5$~M$_{\odot}$. The partition of the horizontal axis is taken from the approximate integer value of the logarithmic function, i.e. masses in the range $\sim$$M_{02} \times 10^n / 3.16$ to $\sim$$M_{02} \times 10^n \times 3.16$ belong to the bin labelled by $n$.
	\ref{fig:histogram-density} Histogram of the clump density distribution, as in the upper panels. Densities were calculated considering the data retrieved by \textsc{Skid} \citep{stadel01}. The mass and volume of a clump used for average density estimates correspond respectively to its total mass and to a sphere with the maximum radius of the clump.
\label{fig:clumps}
}
\end{figure*}

Figure~\ref{fig:clumps} shows the distribution of clump masses at different times (0.5--5~Myr) normalized to the mass of the secondary BH, which is almost constant during this time range in the runs shown in the figure (NAC, CSFSN, and CSFSNBF; see Table~\ref{tab:runs}). In run NAC, especially after 3--5~Myr, there are more than 10 clumps with masses from commensurate to ten times more massive than the secondary BH. As discussed in \citet{fiacconi13} and \citet{roskar15}, clumps with masses comparable to or larger than the mass of the secondary BH can have a dramatic effect on the latter, with direct hits or even repeated softer perturbations. At the same time, Figure~\ref{fig:clumps} shows that most of the clumps have masses about the same or less than the secondary BH, namely close to or below $10^6$~M$_{\odot}$, in the range of typical Giant Molecular Clouds (GMCs).

As previously mentioned, the secondary BH is ejected outside the plane of the disk, above the nominal disk scale height. This effect, however, was strongly influenced by a particular close encounter between the secondary BH and a clump with center of mass less than 3 softenings away at $t = 4.44$~Myr. A proper description of interactions happening at such close ranges should account for their dependency on the detailed structure of the gas clumps. Yet, the numerical prescriptions adopted here are not a comprehensive set of the physical processes happening in the interstellar medium, and the uncertainties on the clump description at this scale may be non-negligible. Furthermore, \citet{delvalle15} noted that for marginally-resolved interactions, decreasing the gravitational softening of the BH can greatly influence the interaction, diminishing the strength of scatterings and potentially disrupting the clump, as the lack of force resolution can render the BH unable to do so.

On the other end, according to the same work, this feature is particularly relevant when the effective density of the BH, defined as $\rho_{\rm BH} = 3\,M_{02} / (4\pi\epsilon_{\rm F}^3)$, is smaller than the density of the clumps. We note that $\rho_{\rm BH}  \sim 1.6 \times 10^8$~amu cm$^{-3}$ is an upper bound for the densities of all clumps at this stage of our simulations (see Figure~\ref{fig:clumps}) and we observe a combined effect of disruption and scattering. Besides, the typical BH-clump scatterings involve distances fairly larger than a couple of softening lengths in our simulations.

Once the ejection happens, the secondary BH finds itself in a lower density background, which decreases the strength of both dynamical friction and disk torques considerably (see Figure~\ref{fig:z_separation}). As a result, in run NAC, the separation between the two BHs is still above 50~pc at 60~Myr, in stark contrast to the results of the adiabatic simulations described in the previous section. The redistribution of disk mass due to the gas inflow can in principle have an important effect on the orbital decay of the secondary BH even in absence of an ejection. Indeed, as shown in Figure~\ref{fig:circular_vel_profiles}, as the CND becomes more centrally concentrated, the radial profile of the circular velocity of the disk (defined as $v_{\rm circ} = \sqrt{{M( < r)}/{r}}$, where $M( < r)$ is the gas mass inside a radius $r$) becomes much steeper in the center over a few disk orbits inside of 50~pc. The secondary BH will inevitably increase its orbital velocity proportionally to the local circular velocity. Therefore, since dynamical friction is inversely proportional to $V_{\rm BH}^2$ in the transonic and supersonic regimes relevant here \citep[see e.g.][]{mayer13}, the drag should be reduced in the limit of a static background \citep{ostriker99}.  The Mach number, which is already larger than unity, also becomes higher as the BH moves faster but the gas remains at similar temperatures (${\sim}10^4$~K) due to fast radiative cooling. This also tends to reduce the drag, which is maximum near the transonic point \citep[see][]{chapon13}. We note, however, that in the simulations the background is not static, and these effects might be less important as the local gas velocity should increase as well as a result of the inflow. The same effect has been invoked to explain the stalling of sub-pc separation binaries in smooth CNDs resulting from galaxy mergers \citep[][]{mayer08}. Figure~\ref{fig:circular_vel_profiles}, though, suggests that this central mass accumulation effect is present to some degree in all runs with cooling, irrespective of other processes modeled in the simulation. In the next section we will assess  its quantitative effect by comparing the orbital decay in the NAC run with that in the runs with star formation.

Next, we performed a set of different simulations featuring combinations of presence and absence of BH accretion, star formation, and SN feedback. We consider now two representative runs (see Table~\ref{tab:runs}): first CSF (Section~\ref{Effects_of_star_formation}) and then CSFSN (Section~\ref{Effects_of_SN}).


\subsubsection{Effects of star formation}\label{Effects_of_star_formation}

GMCs are the sites of star formation. Even before feedback processes are considered, the process of conversion of dissipative gas into a dissipationless stellar component should affect disk stability and fragmentation over time, in particular how tightly bound the clumps can get. This can have important effects on the orbital evolution of the pair of BHs. Indeed, in a scenario of rather efficient star formation, gaseous overdensities form stars rather than continuing to collapse. One expects that, as star formation proceeds, clumps will be less dense and more easily disrupted, reducing the effect of stochastic torques and off-plane scattering. In this section we study the stabilizing effect of star formation in two runs, one without and one with gas accretion onto the two BHs, named, respectively, NACSF and CSF (see Table~\ref{tab:runs}).

\begin{figure}
\includegraphics[width=0.47\textwidth]{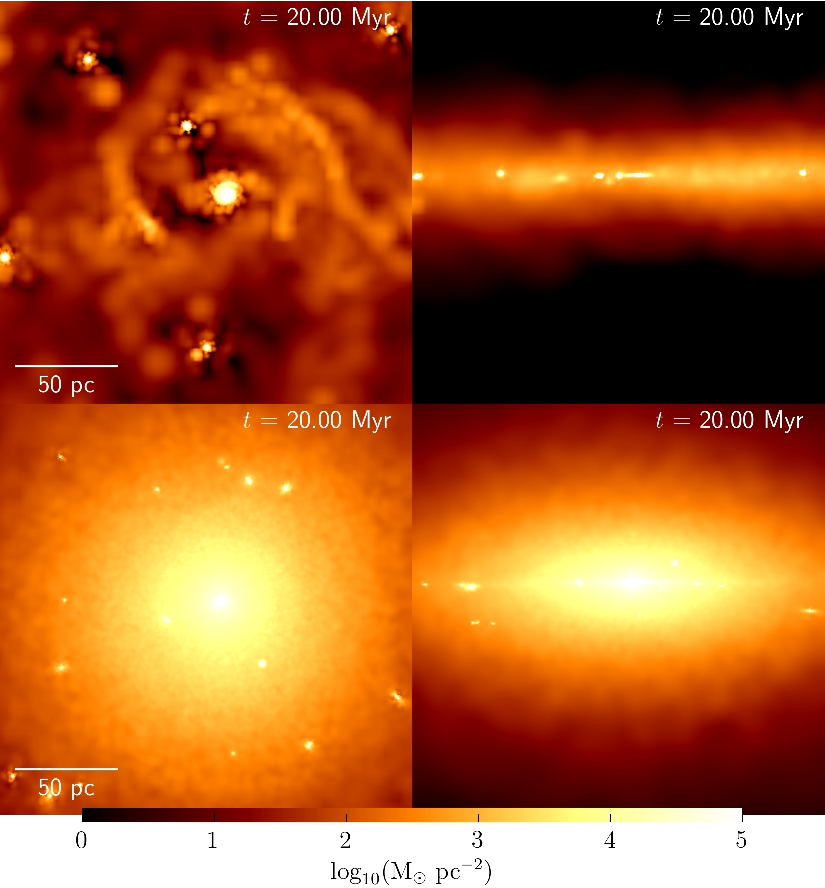}
\caption{Baryonic density maps of runs NAC and NACSF. In the top row we show the face-on (left) and edge-on (right) gas density maps at 20~Myr for run NAC. The bottom row shows the corresponding maps for the stellar density in run NACSF, which is the dominant component of the disk's baryonic content (in this run, the stellar mass surpasses the gaseous mass by a factor larger than 4). Strong star formation leads to a less contrasted density background along the secondary BH trajectory, as the collapsing massive gas clumps give space to stellar overdensities that are less likely to further collapse into a more closely bound structure. By comparatively loosening up the clumps, they become more prone to disruption via encounters, allowing the matter to be redistributed into a smoother disk. A smoother background remits to the adiabatic runs, in the sense that the dynamics of the secondary BH becomes dominated by its interaction with the disk on a local and global scale, whereas strong scatterings from clump encounters play a minor role.
\label{fig:nac_vs_nacsf}
}
\end{figure}

In the simulation NACSF, within 10~Myr the amount of newly formed stars and gas in the average clump are both of the order of a few $10^6$~M$_{\odot}$. Most of the gas has been turned into stars and the CND itself has become a predominantly stellar disk of size similar to the original gaseous one. The rapid conversion of gas into stars not only has suppressed clump formation but also has stabilized the disk globally, leading to a milder evolution of the CND density distribution (Figure~\ref{fig:nac_vs_nacsf}). Yet the difference in the circular velocity profile at 50~pc is only about 10\% (Figure~\ref{fig:circular_vel_profiles}), which suggests that the central mass accumulation effect cannot explain the large difference in the orbital decay time between runs NAC and NACSF (or CSF) seen in Figure~\ref{fig:separation}. Therefore, we believe that it is the global disk stabilization by star formation which reduces the amplitude of stochastic torques and brings the orbital decay back to a more regular path similar to the adiabatic case.
 
Runs with star formation but no feedback are, of course, idealized. In particular, the star formation efficiency is maximized in the absence of feedback. Indeed, as  shown by Table~\ref{tab:mass_ratios}, towards the end of the simulations most of the disk gas has been turned into stars, while observations of CNDs show that the relative weight of gas and stars in the mass budget is at most comparable \citep{medling14}.

Gradual circularization of the BH orbit in the runs NACSF and CSF eventually brings the orbital decay back to the disk-torques driven regime, resulting in a tight pair formation on a $\sim$50~Myr time scale (see Figure~\ref{fig:separation}).

Figure~\ref{fig:z_separation} shows that, while off-plane scattering of the secondary BH occurs, it does so only in the early stage. In Figure~\ref{fig:separation} it is evident that, while initially a delay in the orbital decay occurs relative to the adiabatic runs, the secondary BH ultimately decays on a timescale only a factor of 2 larger relative to the adiabatic runs. 

This is consistent with previous studies showing that the orbital decay in smooth disks is negligibly affected by the relative mass budget of gas and stars as long as the CND has the same structural parameters \citep{dotti07}. This is clearly applicable to the late stages of the star formation runs, as the CND is substantially smoother. This being said, in run CSF we also observe the capture of a gas clump by the secondary BH after its orbit circularizes, at $\sim$45~Myr, which may contribute to accelerating the last phase of the decay (Figure~\ref{fig:separation}) as the secondary acquires a larger effective mass and excites stronger density waves.


\subsubsection{Effects of feedback from supernovae}\label{Effects_of_SN}

As newly formed massive stars explode as SNII, on timescales of 5--6~Myr and longer, we expect blastwaves to develop, heating and pushing gas away, especially in the direction perpendicular to the disk plane, which offers the path of least resistance as the ambient medium has a much lower density. A considerable amount of gas is lost in outflows (at 45~Myr, 30\% of the total remaining amount of gas is found in ejecta in run CSFSN). The effect is evident in Figure~\ref{fig:rho_1e3pc}. Powerful winds develop as the star formation rate, while a factor of 3--4 lower than in the absence of feedback, is still significant compared to the moderate mass of the disk, peaking at $\sim$2.5~M$_{\odot}$~yr$^{-1}$ (Figure~\ref{fig:sfr}). Such star formation rates are on the moderate side for CNDs in merger remnants, based on both observations and simulations \citep{kennicutt98,roskar15} while they correspond to the high end of star formation rates found in the CND of low-redshift Seyfert galaxies \citep{diamond12,esquej14}. As inferred by inspecting Figures~\ref{fig:separation} and \ref{fig:z_separation}, the orbital decay is somewhat erratic at the beginning  (the disk fragments rapidly and initially behaves as in the `cooling and star formation'-only runs until SNII explode) but, after $\sim$10~Myr, no ejection occurs. The secondary BH sinks fast to the center, on a timescale of the same order of the adiabatic runs, in one of the runs (CSFSN), while in the other run (NACSFSN) it wanders significantly at large distances and then enters a slow-decay phase which, however, does not lead to a successful pairing even after 60~Myr. Note that, while the two runs differ for the presence or not of BH accretion (none of them includes BH feedback), the markedly different result is again caused by the stochastic behaviour in the gravo-turbulent CND. In NACSFSN, the inefficient decay is due to the weak drag that the BH suffers once away from the disk midplane, as it finds itself in the low-density envelope produced by SN outflows at heights above 100~pc (see Figure~\ref{fig:rho_1e3pc}). Instead, the other run suggests that, when the BH is not ejected, the drag in the CND is of comparable strength to the cases without feedback, as similar decay timescales are obtained, despite how the CND density distribution and temperature is affected.

Furthermore, here again we observe an eventual tendency of orbital circularization for both the clumps and the BH trajectories (Figure~\ref{fig:separation}). This favours the occurrence of co-orbital encounters between them at late stages of the simulation (after 35~Myr). At this stage, several of those encounters eventually led to the merging between the gas clouds and the BH involved, likely explaining the sudden fast decay in the late stage of the CSFSN run, which includes BH accretion.


\subsubsection{Effects of AGN feedback}

When AGN feedback is implemented, the overall disk morphology is not too different from the run without AGN feedback (Figure~\ref{fig:rho_1e3pc}). However, as it will be explained below, local effects of BH feedback on the interstellar medium play an important role on the orbital decay of the secondary BH.
	
\begin{figure*}
\centering
\includegraphics[width=0.95\textwidth]{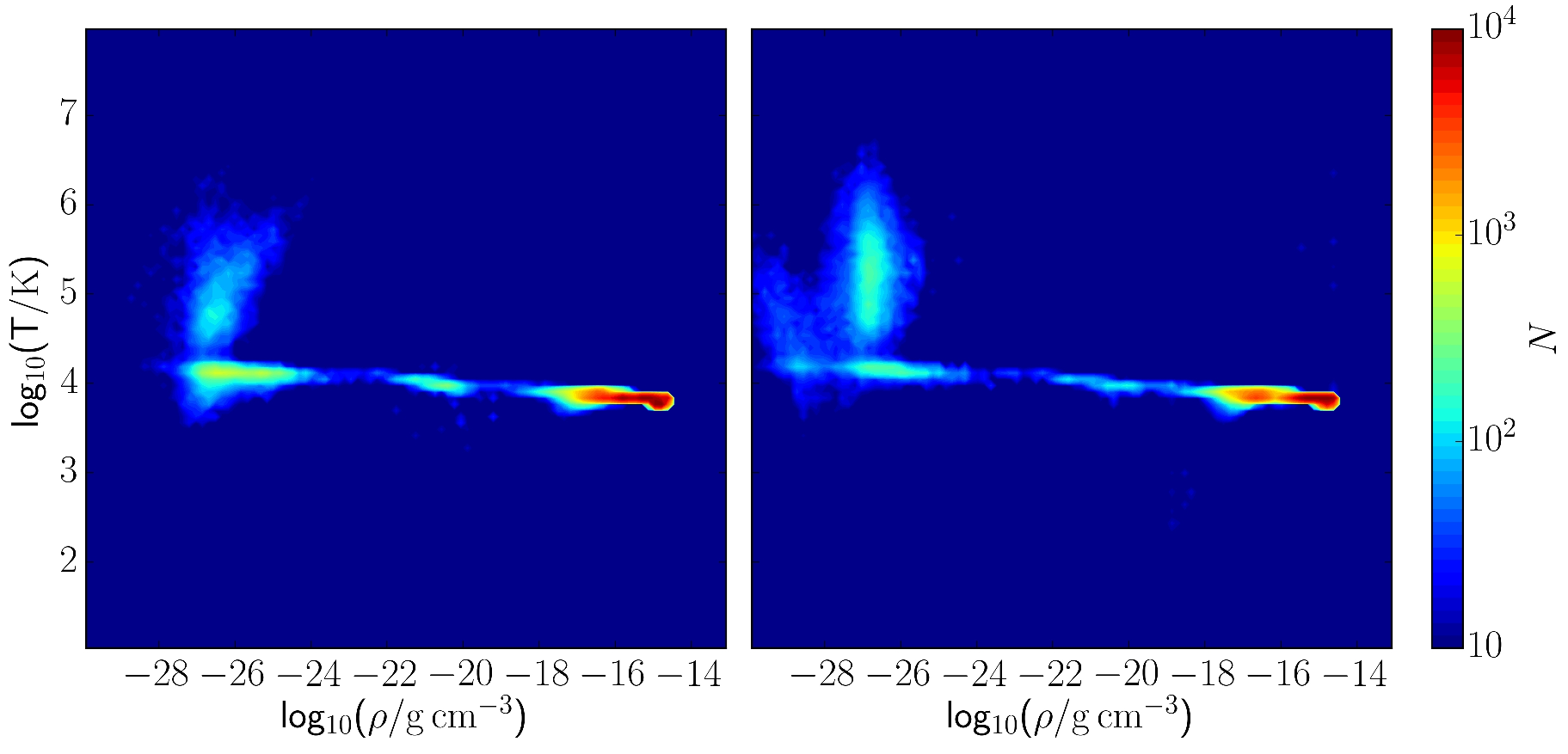}
\caption{ Phase diagrams for runs CSFSN (left) and CSFSNBF (right) at 30~Myr. The color coding shows particle number, so that brighter regions have more particles in the corresponding range of density and temperature. The overall morphology is similar between the two runs but a larger fraction of the gas clearly can reach higher temperatures in the AGN feedback case, as shown by the fact that the diagram is more densely populated above $10^{5.5}$~K. Other differences are likely related to the intrinsic stochasticity of star formation and SN feedback.
\label{fig:phase_diagram}
}
\end{figure*}

Gas outflows from the CND become slightly more prominent than with SNII blastwave feedback alone, developing strong bubbles of very low density and extremely hot gas after several Myr (Figure \ref{fig:phase_diagram}), which reach out to larger distances from the disk midplane compared to the runs with only SN feedback.

After 10~Myr, extensive, very hot bubbles reach out to more than kpc scales, creating a  low-density diffuse  warm/hot envelope (Figure~\ref{fig:rho_1e3pc}) in which dynamical friction onto the secondary BH would be very weak.

\begin{figure}
\includegraphics[width=0.47\textwidth]{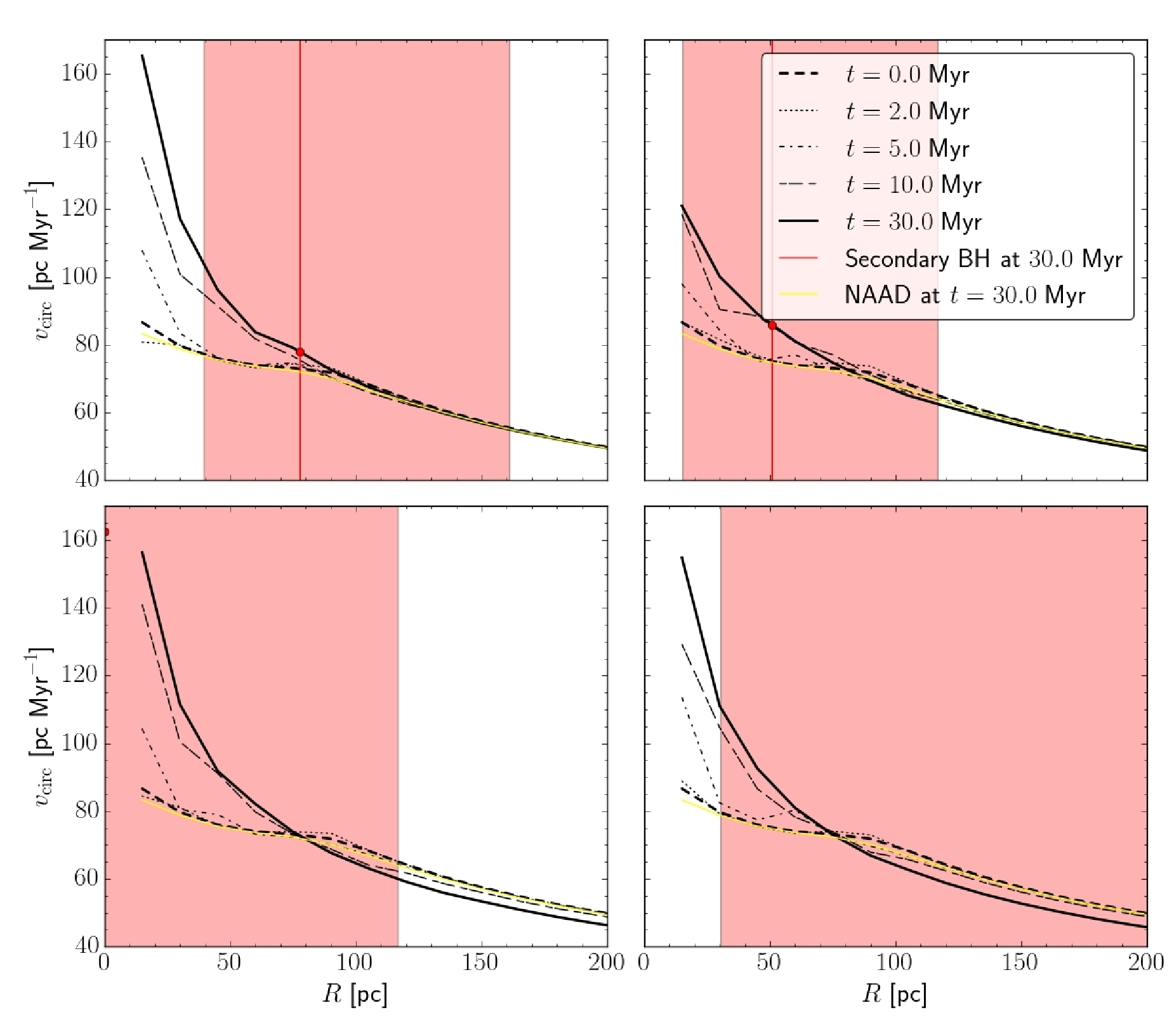}
\caption{Circular velocity profiles centered on the primary BH's position for runs NAC (top-left), CSF (top-right), CSFSN (bottom-left), and CSFSNBF (bottom-right). The red vertical line indicates the radial distance, projected on the disk plane, of the secondary BH from the primary one at $t = 30$~Myr, whereas the shaded region indicates the range of such distances during the first 30~Myr.
\label{fig:circular_vel_profiles}
}
\end{figure}

\begin{figure}
\includegraphics[width=0.5\textwidth]{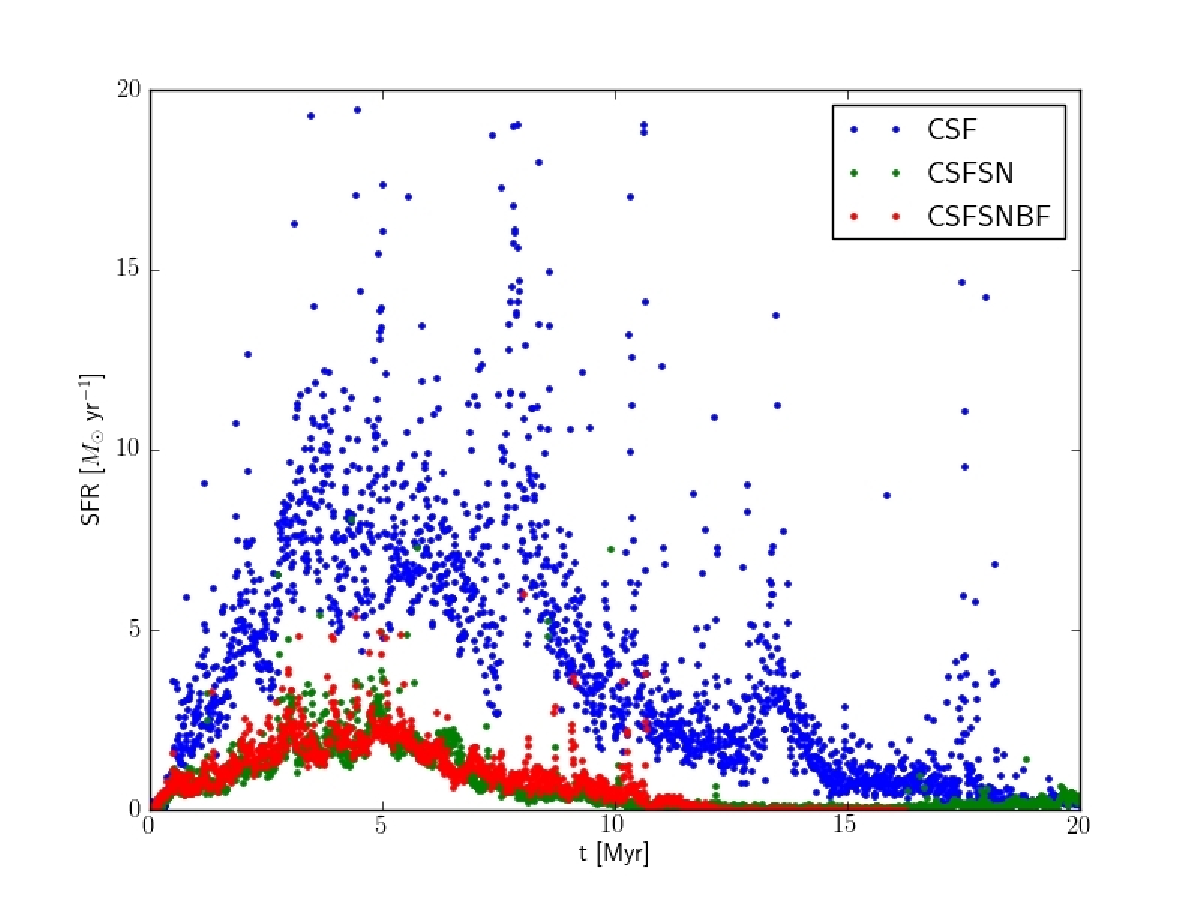}
\caption{Effects of the feedback mechanisms on the star formation rates in the simulations. SN feedback is the key ingredient that suppresses higher star formation rates. The introduction of AGN feedback has a negligible effect in that sense.
\label{fig:sfr}
}
\end{figure}

The outflow, though, is weakly loaded in mass, hence does not have a considerable impact on the mean properties of the CND (see Table \ref{tab:mass_ratios}). In both runs CSFSN and CSFSNBF, the total baryonic mass ejected from the disk (defined as a box of side length 300~pc and 20~pc height centered at the primary BH) at $t$ = 20~Myr is, respectively, $\sim$14\% and $\sim$15\%. This is consistent with the fact that the circular velocity profiles (see Figure~\ref{fig:circular_vel_profiles}) suggest that the gravitational potential wells of the disks themselves are similar between those two runs up to $t = 30$~Myr. Similarly to the runs with SN feedback and no AGN feedback, star formation is essentially truncated after about 10~Myr (Figure~\ref{fig:sfr}), despite the fact that a large reservoir of gas remains (Table~\ref{tab:mass_ratios}). The distribution of clump masses during the initial 5~Myr of the simulation, indeed, does not differ much from that of runs without AGN feedback (see Figure~\ref{fig:clumps}).

These findings may seem in contrast with the expectation that an extra source of energy feedback would suppress further gas densities and subsequently have an impact on star formation rates. On the other hand, it has been reported that the energy output from AGN in a dense disk will have little impact on the disk itself \citep{Roos_et_al_2015}, especially when compared to the energy that couples to the disk coming from stars that go SNe, which is distributed all over the disk. This does not mean that AGN feedback does not have important consequences; it regulates BH growth, and it introduces a new effect that can alter the orbital decay timescale of the secondary BH, as we are going to explain. 

First, AGN feedback has a strong self-regulation action on the mass growth of the secondary BH. Without AGN feedback, the accretion rate is always Eddington-limited as, with a dense gas reservoir, the Bondi rate would often exceed the Eddington rate, while with AGN feedback it leads to Eddington rates of $\lesssim 0.2$. For the primary BH, instead, growth remains Eddington-limited also with AGN feedback, suggesting that, in the very dense nucleus of the CND, AGN feedback is not as effective at self-regulating its mass growth in the way it is implemented in our simulations.

\begin{deluxetable}{lllll}
\tabletypesize{\footnotesize}
\tablewidth{\columnwidth}
\tablecolumns{5}
\tablecaption{Baryonic mass at 20~Myr. By ``mass in a disk'' in this table, we mean the mass within a box centered at the center of mass of the system with 20~pc height and 300~pc side lengths.\label{tab:mass_ratios}
	          $M_{\rm g}$ and $M_{\star}$ are calculated at 10~Myr, whereas $M_{\rm d}$ is the initial disk mass.}
\tablehead{
\colhead{Simulation} & \colhead{$M_{\rm g}/M_{\rm d}$}  & \colhead{$M_{\star}/M_{\rm g}$}    & \colhead{$M_{\rm BH1}/M_{01}$}    & \colhead{$M_{\rm BH2}/M_{02}$}
}
\startdata
NAC		& 0.9766		& 0.0			& 1.0			& 1.0\\
CSF		& 0.1416		& 4.4544        & 1.1081        & 1.0718\\
CSFSN	& 0.7451		& 0.0774        & 1.5569        & 1.3137\\
CSFSNBF	& 0.7423        & 0.0745        & 1.547         & 1.0026
\enddata
\end{deluxetable}

Second, AGN feedback causes what we dub the {\it wake evacuation} effect. For run CBF, which allows to isolate the effects of AGN feedback from SN feedback, it is clearly seen in Figure~\ref{fig:rho_cbf} (top panels). What happens is that, as the secondary BH undergoes a first accretion burst, the bubble produced by AGN feedback primarily evacuates gas behind the secondary BH and trailing with it, which reduces the effect of dynamical friction. The effect plays a role as long as the secondary BH is in the CND. Even after an out-of-plane scattering occurs, it is still noticeable when the secondary BH crosses again the CND along its orbit. The wake evacuation effect explains why, even before any ejection occurs, the orbital decay is really inefficient in runs with AGN feedback (see Figure~\ref{fig:separation} before 10~Myr). Ejections out of the CND plane occur in all our runs with AGN feedback, and in some are the most dramatic in our sample, sending the BH to more than 280~pc from the CND center (e.g. see run CSFSNBF in Figure~\ref{fig:separation}). Wake evacuation effects and ejections combine uniquely in AGN feedback runs to render the orbital decay more inefficient than in all the other runs. Indeed as BHs, due to the slow decay induced by the wake evacuation effect, are ejected when they are still very far from the center of the disk, inevitably their orbital decay is maximally suppressed because they need to lose a larger fraction of their initial orbital energy and orbital angular momentum (Figure~\ref{fig:rho_1e3pc}). Indeed, the secondary BH (in the late stages of run CSFSNBF) keeps orbiting the center at distances between 60 and 280~pc (until we stopped the simulation). In the next section, where we estimate the dynamical friction effect of a large bulge extending out to such large distances, we will discuss the possible timescale for the secondary BH to return to the plane and form a bound BH pair at the center.

\subsection{Estimate of the orbital decay timescale}\label{sec:orbital decay timescale}

Encounters of the secondary BH with clumps in eccentric orbits can eject the BH out of the disk. We have seen several examples of this phenomenon in our simulations. Once outside the CND's midplane, the low-density ambient medium implies very long dynamical-friction timescales. Indeed, our simulations show cases in which the BH pairing appears to stall, in the sense that the separation evolves little or even increases in some cases after tens of Myr. However, if we had included an extended stellar bulge in our simulations, this would have exerted dynamical friction onto the secondary BH even well outside the CND. Note that the stellar (Plummer) bulge already included in our simulation setup only represents a small fraction of the total extent of the bulge of a massive spiral or galaxy merger remnant, its scale radius being more typical of a galactic stellar nucleus (50~pc). This choice was dictated by reasons of computational efficiency.

We can thus estimate what effect would an extended bulge have on the orbital decay timescales down to the formation of a hard BH binary\footnote{When the BHs inter-separation remains under 5~pc after several orbits.}. As we show below, we find an upper limit of $\sim$$3 \times 10^8$~yr for the orbital decay to negligibly small separations, where emission from gravitational waves would take place,  if we consider that the decay is driven by dynamical friction from a fictitious \citet{hern90} bulge. We further have to assume that the BH is moving in a circular orbit and that no subsequent ejections occur. This is a rather long timescale, a fraction of a Gyr, but it is only a factor of 3 longer than the timescale obtained in the merger simulations of \citet{roskar15}, where extended stellar and dark matter distributions were present by construction, but in which many of the processes studied in this paper, in particular BH accretion and feedback, were not taken into account. The likelihood of the occurrence of further ejections is clearly related to how clumpy the disk will be by the time it recaptures the secondary BH, which depends upon the lifespan of massive clumps and on the timescale in which the nuclear region of a galaxy is accreting gas and replenishing its gas reservoir (or not, if the effect of feedbacks is very strong). Since in the simulations we consider a disk in isolation without gas inflows from the outer regions of the host galaxy, clumps inevitably become more rare after several tens of Myr.

We resort to the Hernquist analytical model in an attempt to consider an extended bulge for our estimate. By taking into account the effects of dynamical friction in a fictitious stellar Hernquist bulge of scale radius $a$ and total mass $M_{\rm b}$, the corresponding timescale for a body of mass $M$ in circular motion inside such a bulge to inspiral from an initial radial distance $r_{\rm i}$ to a final one $r_{\rm f}$ is given by

\begin{equation}
t_{\rm df} = \frac{1}{4 \xi \ln \Lambda} \sqrt{\frac{a^3}{G M_{\rm b}}} \frac{M_{\rm b}}{M} \int_{\chi_{\rm f}}^{\chi_{\rm i}} \frac{\chi^{3/2}(\chi + 3)}{\chi + 1} d\chi,
\label{eq:decay timescale results}
\end{equation}

\noindent where $\xi$ is a correction factor of order unity, $\ln \Lambda$ is the Coulomb logarithm, and $\chi = r/a$ is the ratio between the radial coordinate of the orbiting mass, $r$, and the bulge's scale radius, $a$. Under the assumption that $\chi_{\rm i} \ll 1$ (i.e. $r_{\rm i} \ll a$), the integral in Equation (\ref{eq:decay timescale results}) can be approximated by $6 \chi^{5/2}/5$, and we can re-write the equation as

\begin{align}
t_{\rm df} &= 372\ \xi^{-1}\ \textrm{Myr} \left( \frac{\ln \Lambda}{5} \right)^{-1} \left( \frac{M_{\rm b}}{10^{10} {\rm M}_{\odot}} \right)^{1/2} \nonumber\\
       &\times \left( \frac{M}{5 \times 10^5 {\rm M}_{\odot}} \right)^{-1} \left( \frac{a}{750 \, \textrm{pc}} \right)^{3/2} \left( \frac{r_{\rm i}/a}{0.4} \right)^{5/2}.
\label{eq:decay time results}
\end{align}

For the parameters of the bulge, we consider $a = 750$~pc and $M_{\rm b} = 10^{10}$~M$_{\odot}$. For the secondary BH, we take $M = 5 \times 10^5$~M$_{\odot}$ (the initial mass of the BH), $r_{\rm i} = 300$~pc, and $r_{\rm f} = 0$~pc. The choice of $r_{\rm i}$ is motivated by the fact that the secondary BH can be almost this far from the center near apocenter in some of the runs, e.g. those with AGN feedback, in which strong ejections occur. Taking $\ln \Lambda = 5$ \citep[see][]{binney08} and $\xi = 1$, we find (using Equation \ref{eq:decay timescale results})

\begin{equation}
t_{\rm df} \simeq 0.3 \; \textrm{Gyr}.
\end{equation}

Since in many of our simulations ejections occur to smaller distances than 300 pc, the upper limit on the orbital decay timescale will change accordingly. For instance, a large fraction of runs have secondary BHs ejected to 100--150~pc (see Figure~\ref{fig:separation}), which would yield an expected BH hard binary formation timescale of the order of 2--$7 \times 10^7$~yr \citep[of the order of what found in][]{roskar15}. However, for the AGN feedback runs the 300 pc limit is representative as ejections occur to comparable distances. In the next section we comment on the implications of such rather long timescales.

\section{Summary and Discussion}\label{sec:conclusions}

The suite of CND simulations with embedded massive BH pairs presented in this paper represents the most comprehensive study so far of all the physical processes that can affect the orbital decay of massive BHs from 100-pc separations to separations of fractions of pc. Looking at the essential picture emerging from the simulations, it is clear that the orbital decay timescale is generally increased relative to the adiabatic runs, but also that it can span a really wide range, from 20~Myr to an estimated maximum of more than 300~Myr (given that no further ejections occur).

{\it First}, our simulations confirm that, as long as radiative cooling is included, the CND becomes highly clumpy due to fragmentation, with masses of clumps as large as $10^7$~M$_{\odot}$, which corresponds to the largest GMC masses, in essentially all the runs despite the inclusion of other processes such as feedback (Figure~\ref{fig:clumps}). Indeed,  our clumps are generally more compact and much denser than the typical GMC in the \emph{disk} of the Milky Way.  Despite the richer physics of the simulations presented here, the sizes and densities of the clumps are similar to those found in \citet{fiacconi13}, for which it was shown that a good match exists with clouds in the Galactic Center region \citep{oka01}, which, albeit less extreme, represents an environment with high gas densities and stronger stellar radiation background more akin to what one expects in the nucleus of a merger remnant, which is what our CND model should capture. We repeated the same comparison here (Figure~\ref{fig:d_vs_m}). Note that the radii adopted for our clumps are the outermost radii for which bound material is found by our group finder \textsc{Skid}. While other definitions could be given for the size of a clump, we chose this because it is consistent with the analysis in \citet{oka01},  in the sense that they only show results  for clouds with diameter of about 4 pc or larger (due to resolution limits). Clumps and overdense spiral arms introduce a {\it stochastic orbital decay} regime, leading to perturbations as well as direct hits to the secondary BH, which almost invariably scatter it out of the CND plane. Once outside the CND plane, dynamical friction is considerably reduced, increasing the orbital decay timescale by at least an order of magnitude. This is the dominant effect in all runs, confirming the results of \citet{fiacconi13} and \citet{roskar15}.

{\it Second}, as the CND becomes unstable, mass flows inward, steepening the CND density and circular velocity profile relative to the adiabatic case. This steepening, however, has relatively little effect on the orbital decay of the secondary BH, compared to, e.g., clumpiness.

{\it Third}, star formation and feedback processes decrease the gas content of the CND. Feedback leads to outflows that produce a diffuse hot envelope around the CND. However, globally outflows do not have a measurable impact on the orbital decay as most of the mass stays in the CND midplane. Likewise, the orbital decay in a mostly stellar CND is almost identical to that in a gaseous CND as long as densities
are comparable, which also confirms previous results \citep[][]{dotti07}.

{\it Fourth}, the warmer medium produced by feedback processes in the CND, though, also concurs to reduce the strength of the drag, especially in the dynamical friction phase \citep[see also][]{park17}. The extreme manifestation of this is the {\it wake evacuation effect} in AGN feedback runs, in which a hot bubble driven by feedback around the secondary BH lowers drastically the gas density behind the secondary, where the dynamical friction wake should form, stifling orbital decay.
 
Momentum feedback via jets, for example, would act perpendicular to the disk plane, hence would not generate an evacuation of the wake in the disk midplane, which is what is relevant to affect dynamical friction and disk torques. However, radiation feedback would still have an effect in the midplane even if it happens anisotropically rather than isotropically as we assume here with Bondi. The reason is that the stronger drag from the wake will come from the densest material just behind the BH, at a few pc from it, namely close to the sphere of influence. This scale is small enough that the overall geometry of the radiation front and its relation to the vertical structure of the disk should not matter much. That is, even if the effect would be less dramatic on a large scale (${\sim}10$~pc), with only the inner part of the wake evacuated, that would be enough to have an impact on the drag suffered by the BH.

Regarding the first effect, we notice that there is a subset of runs in which clumps can actually have a positive feedback, accelerating the orbital decay time. For example, in run CSFSN a clump scattered the BH towards the center, promoting a shorter decay. We also observed post-orbit-circularization co-orbital encounters between clumps and the BH, much more common at later stages of the simulations (depending on the run, but roughly after 30~Myr). These can also aid decay as the secondary BH accretes mass efficiently from a clump or even merges with it, reaching a higher mass that amplifies the negative torque responsible for its decay. Some gas can remain in a sub-disk around the secondary BH, which can also have a positive effect on the decay as it torques directly onto the sub-disk formed around the primary BH in the last stage of the decay (yet this effect will have to be studied better with higher-resolution simulations delving much deeper in the sub-pc regime of the decay).

Our simulations do have a number of caveats. To mention a few, the radiative cooling model used in this work is less sophisticated than that employed by \citet{roskar15}, as it lacks to account for metal lines above $T  = 10^4$~K, molecular and dust-gas cooling, as well as self-shielding at high densities/opacity. However, high-temperature metal cooling is less important in isolated CND simulations as there are no shocks as in a galaxy merger, yet it could affect SN and AGN-driven hot outflows by cooling gas faster out of the bubbles. Nevertheless, since outflows have little effect on the orbital decay, we do not expect this to be a major limitation for our purpose. Self-shielding would reduce cooling at the highest densities, but since we also do not include molecular cooling or cooling via dust-gas collisions, which dominate at the highest densities, the new result should be similar to what we model here. Indeed the masses and characteristic densities of clumps in our simulations are similar to those in \citet{roskar15}. In the literature there are also newer prescriptions for BH accretion and feedback which take into account angular momentum loss by torques or effective viscosity in accretion disks, which is a more realistic approach relative to Bondi accretion. Some of these recipes have been shown to lead to smaller accretion rates relative to Bondi. However, since the secondary BH accretes already at markedly sub-Eddington rates with AGN feedback, such recipes should have little impact on our results (they would on the growth of the primary BH, though, but this has no relevance for orbital decay). Our systems were considered in isolation, therefore there is room to address the effects of galactic gas replenishing the nuclear disk. And, of course, a future, even larger suite of runs should explore varying initial conditions parameters such as the mass of the disk, the initial mass of the secondary BH or its orbital inclination with respect to the disk. The latter approach was used in some of the previous works appeared in the literature \citep[e.g.][]{fiacconi13,delvalle15}.

The BH masses considered in this work are within the detection window of eLISA \citep{amaro-seoane2013}, hence the orbital decay timescales that we estimate are of high interest. BH tight binary formation timescales ranging from a few times $10^7$ to a few times $10^8$~yr are in contrast with the once prevailing idea that orbital decay should be accelerated in gaseous backgrounds as opposed to stellar backgrounds \citep{mayer07,mayer13}. A similar delay effect due to a clumpy interstellar medium has been studied recently also at larger galactic scales \citep{tamburello2016}. However, while in the latter case disk fragmentation into clumps might be important only in the most massive galaxies at high redshift, in the dense nuclear regions studied here this behaviour appears to be very generic, being now supported by several calculations done with different setups and  simulation techniques \citep{fiacconi13,roskar15,delvalle15}. Here we have enriched the scenario by including new effects induced by feedback processes, most notably wake evacuation, that can delay the decay even in absence of ejections. On the other hand, \citet{khan2016} have recently shown, using cosmological simulations, that the orbital decay can proceed really fast, in a few Myr, in the dense cores of galaxies at $z > 2$ when such cores are devoid of gas and 3-body encounters with stars govern the decay process. Putting these two results together, it would appear that, while gas dissipation is needed to create dense cores in galaxies, a fast decay requires at the same time that such gas dissipation stops being important when the BH pair has been delivered to the nucleus after the merger. This is mimicked in an idealized way by our runs with only star formation and no feedback effects in that they maintain a prolonged phase where dissipative gas physics and a clumpy interstellar medium exist. And, indeed, in the latter case the orbital decay is only marginally longer than in the adiabatic, non-dissipative gas cases. Also, it would be shortened further by the concurrent effect of 3-body interactions, that we do not simulate here. 

This suggests that there might be at least two regimes to take into account. Namely, BH mergers might be less efficient in gas-rich galaxies, which are typically disk-dominated at all radii, and be more efficient in galaxies that have already built a substantial bulge, in which the nuclear gas content is expected to be low. It remains to be understood what this implies in terms of the eLISA discovery space, since one can imagine a variety of galaxy hosts, more or less affected by dissipation, which could harbour BHs with masses in the range relevant for gravitational wave detection. Nevertheless, our results show that future studies making predictions for massive BH merger rates must somehow take into account the nature of galaxy hosts and the fact that there might be different regimes of orbital decay depending on the properties of such hosts.

On the other side of the coin, a longer decay time in the 100-pc to pc-scale separation is interesting for observations of tight double AGN. Studies of dual AGN activity in galaxy mergers \citep{wassenhove12, capelo16b} have already shown that strong dual AGN activity occurs at relatively small separations ($\lesssim 1$~kpc). As many of our BH pairs will remain at $\sim$100-pc separations for fractions of a Gyr, this prompts searches for double AGN at small separations in gas-rich, disk-dominated galaxies with small bulges. Yet, as the secondary BH, once ejected, is harboured in the low density, hot envelope around the CND, accretion is suppressed, and so would dual AGN activity. Possible dual AGN activity in this regime will have to be studied in future work.

\section{Acknowledgements}
This work is supported by the STARFORM Sinergia Project funded by the Swiss National Science Foundation. Part of the analysis was performed using the open source \textsc{Pynbody} package \citep{pontzen13}. RSL thanks Davide Fiacconi for his help. PRC acknowledges support by the Tomalla Foundation.

\bibliography{references}

\begin{appendix}

\section{Additional figures}

\begin{figure}
	\centering
	\includegraphics[width=0.45\textwidth]{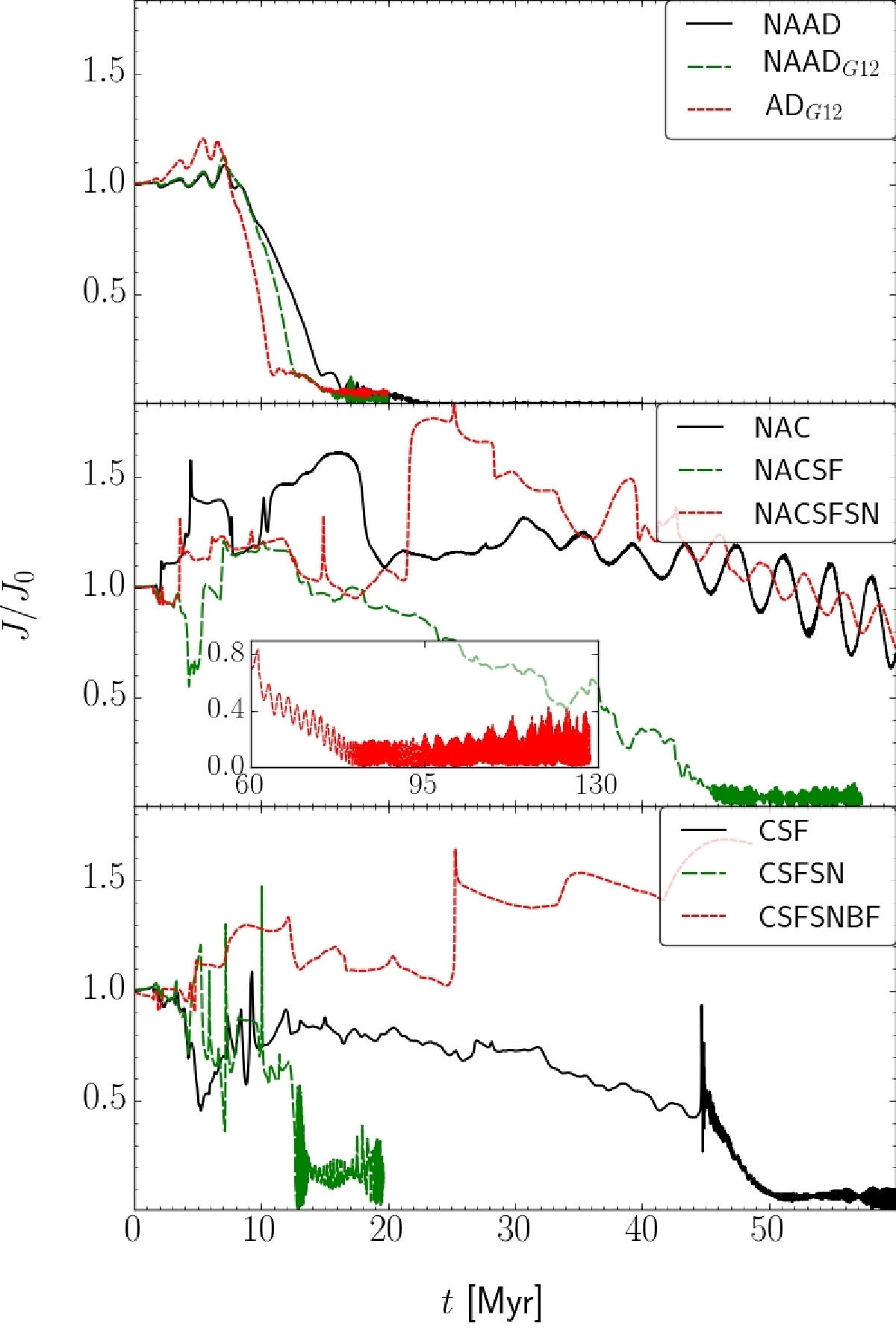}
	\caption{Angular momentum of the secondary BH with respect to the center of mass of the entire system (all particles) for different runs (refer to Table~\ref{tab:runs}). We have to bear in mind that ultimately we want to address the problem of the angular momentum of the system composed by the two BHs (in whichever external forces field they may lie on). This is \emph{not} what is shown in this plot. Since the primary BH is very massive and lies close to the center of the whole system, showing the right quantity would only introduce the noise caused by the primary BH's motion. Here, we opted to highlight the interplay of the secondary BH with the mass field.
		\label{fig:angular_momenta}
	}
\end{figure}

\begin{figure}
	\centering
	\includegraphics[width=0.7\textwidth]{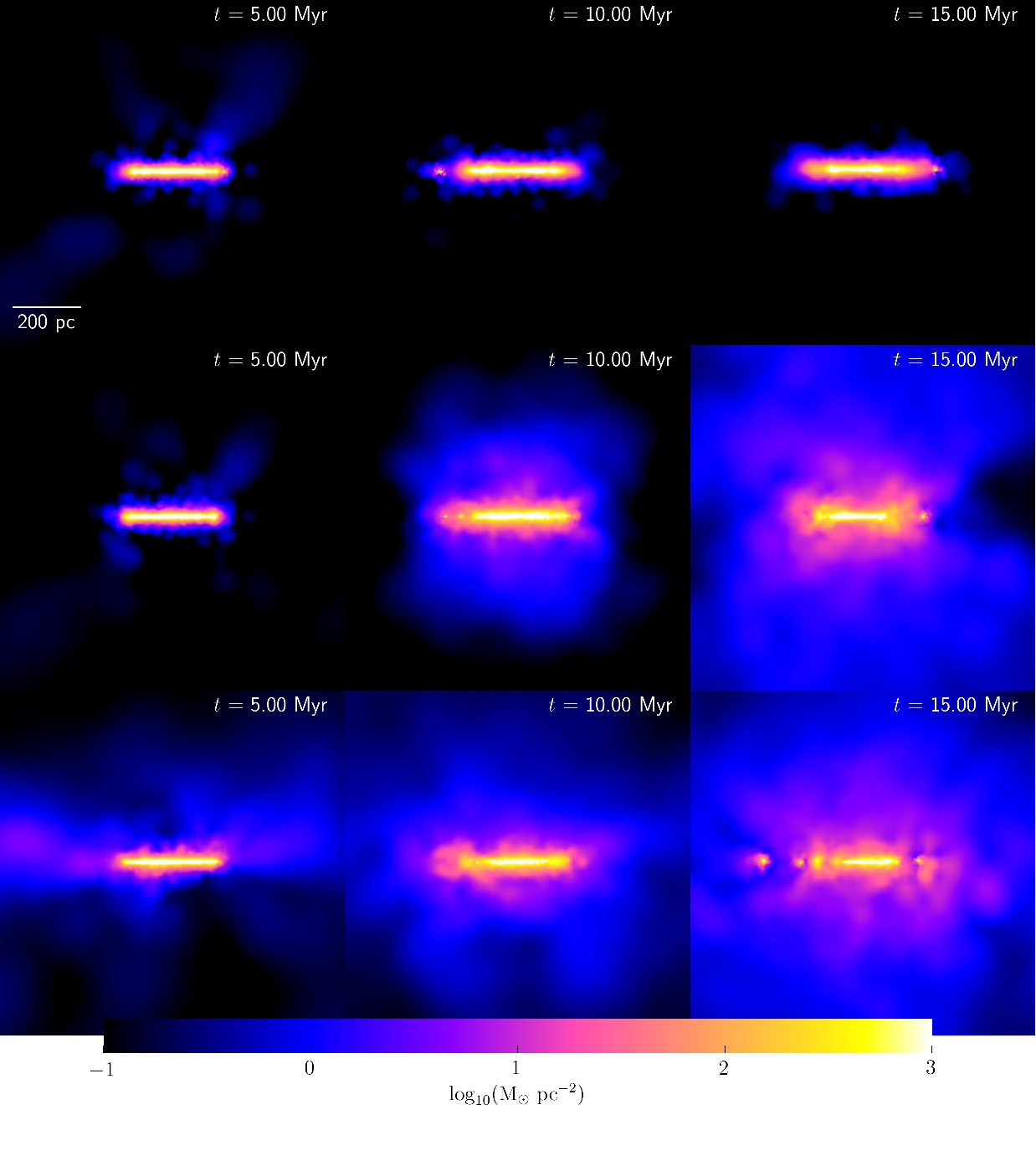}
	\caption{Snapshots of projected gas density (edge-on view) with 1~kpc wide box. In each row, from left to right, there are three different snapshots every 5~Myr of a given simulation. The runs are, from top to bottom, respectively, NAC, CSFSN, and CSFSNBF.
		\label{fig:rho_1e3pc}
}
\end{figure}

\begin{figure}
	\centering
	\includegraphics[width=0.7\textwidth]{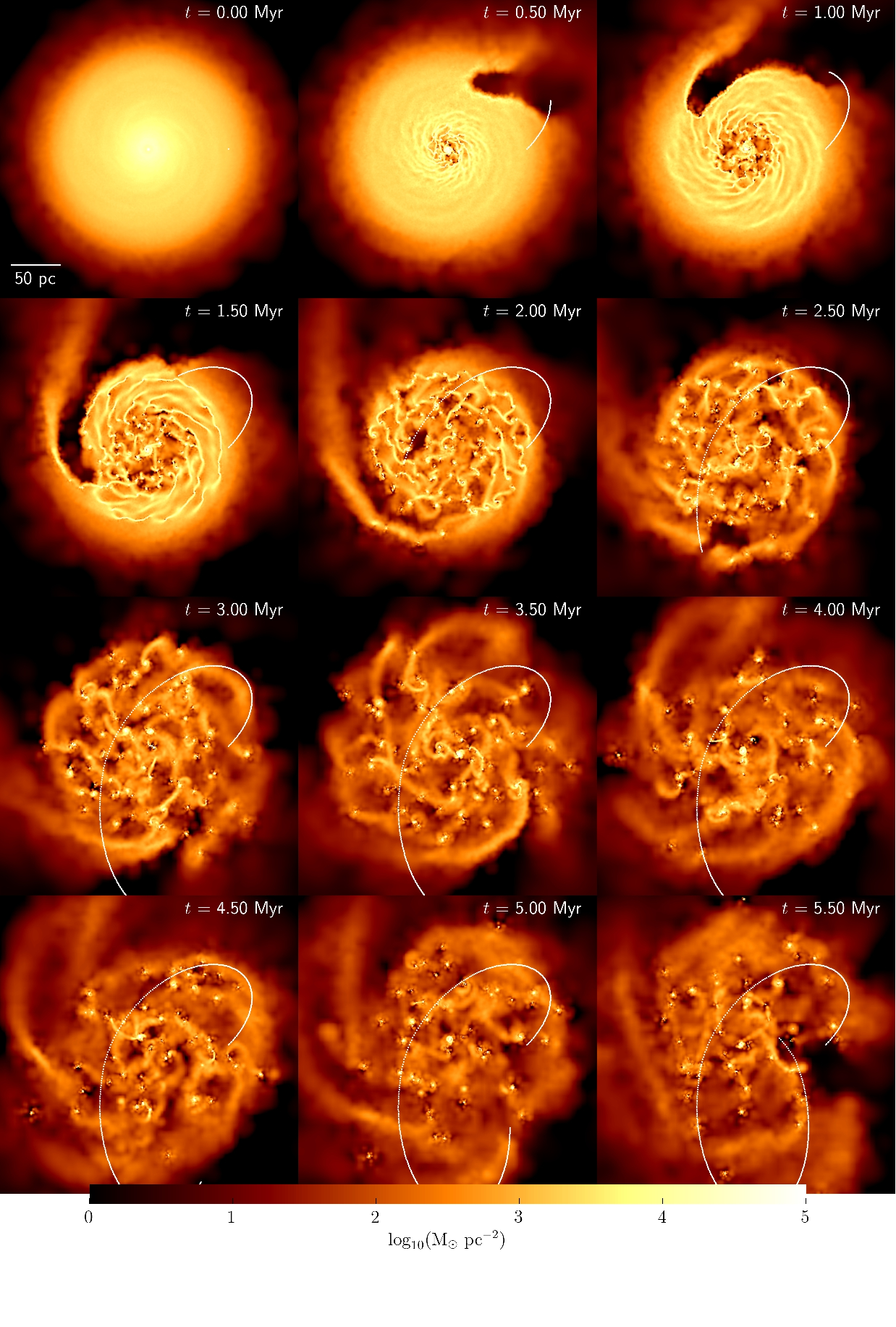}
	\caption{Density snapshots for run CBF. 
			\label{fig:rho_cbf}
}
\end{figure}

\begin{figure}
	\centering
	\includegraphics[width=0.5\textwidth]{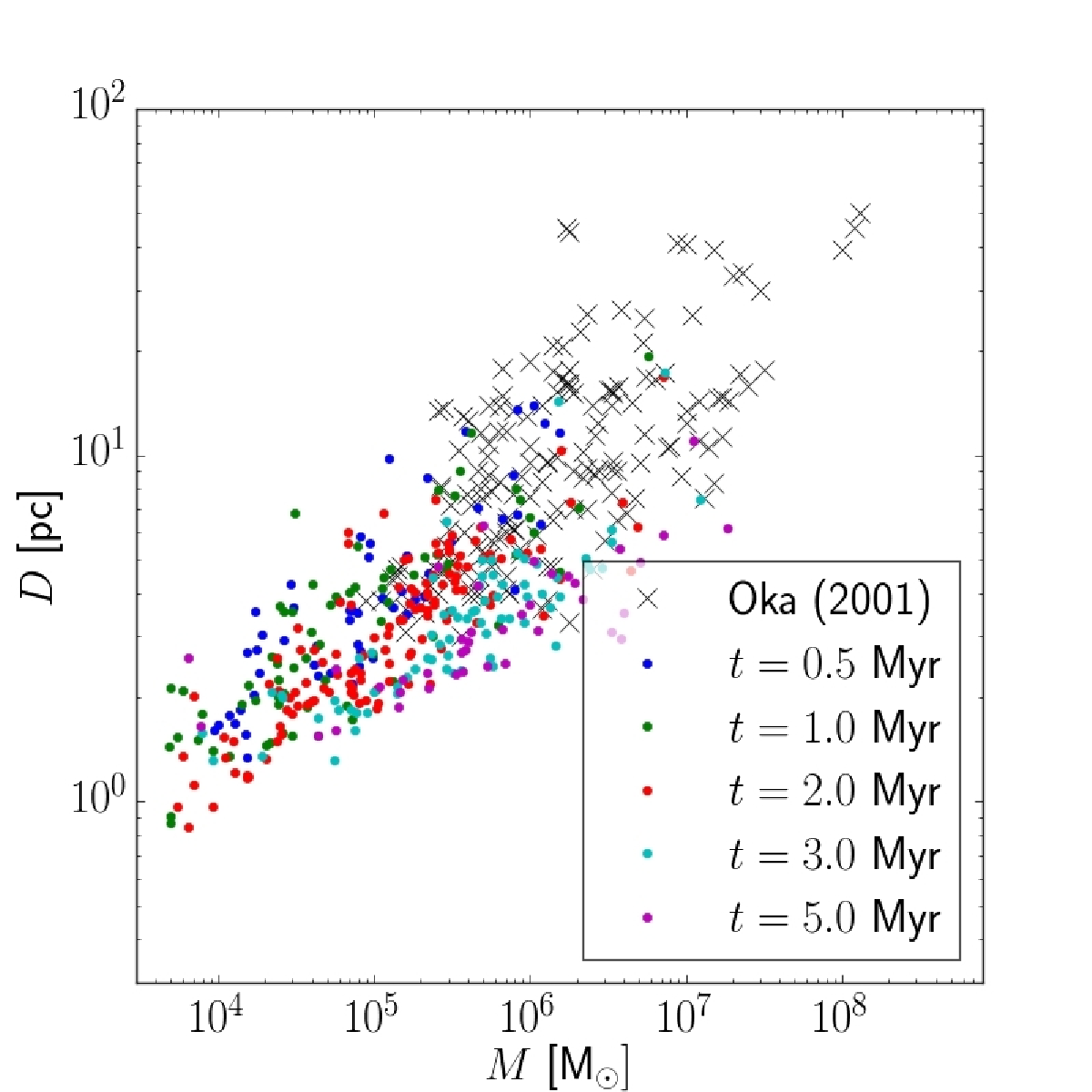}
	\caption{Diameter $D$ (or size $S$, from \citet{oka01}) vs mass $M$ for run CSFSNBF. The clumps from the simulations are in agreement with the available data from the observations of molecular clouds in the Galactic Center. The Galactic Center, however, may represent a less extreme environment than a CND formed after a gas-rich galaxy merger \citep{fiacconi13}. Smaller clouds are difficult to detect, what might justify the lack of observational data for clumps of sizes not larger than a few pc.
		\label{fig:d_vs_m}
	}
\end{figure}

\end{appendix}

\end{document}